\documentclass[noinfobox]{mdolab-article}

\usepackage[sort&compress,numbers]{natbib}

\usepackage{cleveref}
\usepackage{bm}
\usepackage{algorithmic}
\usepackage{algorithm}
\usepackage{caption}
\usepackage{subcaption}
\usepackage{transparent}
\usepackage{multirow}
\usepackage{booktabs}
\usepackage{arydshln}

\crefname{equation}{Eq.}{Eqs.}
\Crefname{equation}{Eq.}{Eqs.}
\crefname{figure}{Fig.}{Figs.}
\Crefname{figure}{Fig.}{Figs.}

\graphicspath{ {../figures/} }
\DeclareGraphicsExtensions{.pdf,.png}

\title{Accelerating Bayesian Inference \\
via Multi-Fidelity Transport Map Coupling}
\author{Sanjan C. Muchandimath~\footnote{PhD Candidate, Aerospace Engineering, AIAA student member.}, Joaquim R. R. A. Martins~\footnote{Pauline M. Sherman Collegiate Professor, Aerospace Engineering, AIAA fellow.}, and, Alex A. Gorodetsky~\footnote{Associate Professor, Aerospace Engineering, AIAA member.}}
\affil{University of Michigan, Ann Arbor, MI, 48109}


\begin{document}

\maketitle

\begin{abstract}
    Mathematical models in computational physics contain uncertain parameters that impact prediction accuracy. In turbulence modeling, this challenge is especially significant: Reynolds averaged Navier--Stokes (RANS) models, such as the Spalart--Allmaras (SA) model, are widely used for their speed and robustness but often suffer from inaccuracies and associated uncertainties due to imperfect model parameters. Reliable quantification of these uncertainties is becoming increasingly important in aircraft certification by analysis, where predictive credibility is critical. Bayesian inference provides a framework to estimate these parameters and quantify output uncertainty, but traditional methods are prohibitively expensive, especially when relying on high-fidelity simulations. We address the challenge of expensive Bayesian parameter estimation by developing a multi-fidelity framework that combines Markov chain Monte Carlo (MCMC) methods with multilevel Monte Carlo (MLMC) estimators to efficiently solve inverse problems. The MLMC approach requires correlated samples across different fidelity levels, achieved through a novel transport map-based coupling algorithm. We demonstrate a 50\% reduction in inference cost compared to traditional single-fidelity methods on the challenging NACA~0012 airfoil at high angles of attack near stall, while delivering realistic uncertainty bounds for model predictions in complex separated flow regimes. These results demonstrate that multi-fidelity approaches significantly improve turbulence parameter calibration, paving the way for more accurate and efficient aircraft certification by analysis.
\end{abstract}

\section*{Nomenclature}

{\renewcommand\arraystretch{1.0}
\noindent
\begin{tabular}{@{} l l @{\hskip 3em} l l @{}}
$J_T$ & Jacobian of the transport map & $\mathcal{K}$ & Markov kernel \\
$N$ & number of samples & $Q$ & quantity of interest (e.g., lift) \\
$T$ & transport map & $d$ & dimension of the parameter set \\
$f$ & forward model function & $q$ & proposal distribution \\
$\bm{x}$ & simulation setup vector (boundary conditions, mesh) & $\bm{y}$ & output vector of the forward model \\
$\bm{y}_{\mathrm d}$ & observed data vector & $\alpha$ & acceptance ratio in the MCMC algorithm \\
$\bm{\epsilon}$ & noise vector in the data & $\gamma$ & joint probability measure between $\mu$ and $\nu$ \\
$\mu$ & probability measure of $\pi$ & $\pi$ & posterior density \\
$\rho$ & Pearson correlation coefficient & $\bm{\Sigma}$ & covariance matrix \\
$\bm{\theta}$ & model parameters & $\bm{\phi}$ & transport map parameters \\
\end{tabular}
}

\vspace{1em}
\noindent
\textbf{Subscripts}\\
\begin{tabular}{@{} l l @{\hskip 3em} l l @{}}
$\ell$ & $\ell$'th fidelity level & $r$ & reference \\
\end{tabular}

\vspace{1em}
\noindent
\textbf{Superscripts}\\
\begin{tabular}{@{} l l @{\hskip 3em} l l @{}}
$i$ & current iteration & $^{*}$ & proposal \\
\end{tabular}

\section{Introduction}

Parameter uncertainty in mathematical models of physical systems is a significant source of error in computational predictions. Understanding how these uncertainties propagate to the outputs of interest is crucial for decision-making in engineering design and certification processes, particularly in aerospace applications where safety-critical decisions often rely on accurate predictions of complex flow phenomena. Reynolds averaged Navier--Stokes (RANS) turbulence models exemplify this challenge, where empirical closure coefficients calibrated against limited experimental data lead to uncertainty in the model predictions.

RANS models such as $k-\epsilon$~\cite{launderNUMERICALCOMPUTATIONTURBULENT1983}, $k-\omega$~\cite{wilcoxFormulationKwTurbulence2008}, and Spalart--Allmaras (SA)~\cite{spalartOneequationTurbulenceModel} follow the Boussinesq hypothesis and come under the class of linear eddy-viscosity models. These models map the unknown turbulent stresses to the strain rate of the mean flow through a linear function involving empirical constants (model parameters). These parameters are typically calibrated against simple canonical flows, yielding suboptimal performance for complex configurations involving adverse pressure gradients and separation —-- flow regimes critical to aircraft performance~\cite{schaeferUncertaintyQuantificationSensitivity2017}. This parameter uncertainty, combined with model form errors~\cite{duraisamyTurbulenceModelingAge2019,xiaoQuantificationModelUncertainty2019}, represents epistemic uncertainty that significantly impacts predictive accuracy and poses challenges for aircraft certification by analysis processes~\cite{maueryGuideAircraftCertification,slotnickCFDVision2030}.

Addressing this parametric uncertainty requires systematic calibration of the model parameters against experimental data, along with quantification of the resulting uncertainty in the model predictions. A class of approaches propose to use supervised learning techniques, where the model parameters are treated as hyper-parameters and are optimized using a training set of data~\cite{daronchSensitivityCalibrationTurbulence2020,ShanDataAssimilationMachineLearning}. But these point-estimate methods do not provide any information about the uncertainty in the model parameters or the predictions, often requiring large amounts of data to obtain accurate estimates. In contrast, Bayesian inference approaches treat the model parameters as random variables and uses Bayes' theorem to update the prior distribution of the parameters based on observed data. This approach not only provides a posterior distribution of the model parameters, but also works remarkably well with small data sets~\cite{robertMonteCarloStatistical2004}.

Recognizing the advantage of Bayesian methods, several researchers have explored Bayesian turbulence model calibration.~\citet{papadimitriouBayesianUncertaintyQuantification2015} approximated the posterior distribution of the model parameters with a Gaussian and optimized the distribution parameters using a maximum likelihood approach. They employed adjoints to compute both the maximum a-posteriori (MAP) estimate and the Hessian of the likelihood, enabling analytical variance estimation.~\citet{cheungBayesianUncertaintyAnalysis2011} focused on the calibration of the SA model coefficients for incompressible, boundary layer flows. They used Markov chain Monte Carlo (MCMC) methods~\cite{hastingsMonteCarloSampling1970} to get samples from the parameter posterior distribution. They also compared competing models by incorporating additive and multiplicative terms in the likelihood formulation. Similarly,~\citet{edelingBayesianEstimatesParameter2014} calibrated the Launder-Sharma turbulence model against 13 different experimental data sets independently. ~\citet{rayBayesianParameterEstimation2016} also used MCMC methods to calibrate parameters in the $k-\epsilon$ model for a jet-in-cross flow. More recently,~\citet{maruyamaDatadrivenBayesianInference2021,daronchSensitivityCalibrationTurbulence2020} performed parameter calibration for more complex configurations such as airfoils, wings and full aircraft. The computational expense of MCMC methods makes direct application to high-fidelity CFD completely impractical --- a single RANS simulation may require hours while MCMC demands tens of thousands of forward evaluations. Consequently, all these works resort to computationally cheap \textit{data-driven surrogate models} as proxies for the expensive forward model or resort to approximate Bayesian techniques~\cite{doroninaturbulenceABC} to enable practical parameter inference.

Reliance on surrogate models creates a fundamental trade-off between computational cost and accuracy. While surrogate models are essential for making Bayesian inference feasible, their inherent approximation error introduces bias when used as a substitute for the true model. This bias arises because surrogates only approximate the true input-output relationship, leading to discrepancies between the true and surrogate model posteriors. Improving surrogate accuracy can reduce this bias, but requires additional expensive simulations, counteracting the original goal of using the surrogates~\cite{rayBayesianParameterEstimation2016}. 

Multi-fidelity methods have emerged as a promising alternative to surrogate-based methods. These methods strategically combine a limited number of high-fidelity model evaluations with a larger number of low-fidelity model evaluations through multilevel Monte Carlo (MLMC) estimators~\cite{gilesMultilevelMonteCarlo2015} to achieve accurate estimates at a reduced computational cost. The key to the success of multi-fidelity methods is to ensure high correlation between the low-fidelity and high-fidelity outputs, which leads to significant variance reduction in the estimators. However, existing multi-fidelity methods for inverse problems~\cite{dodwellHierarchicalMultilevelMarkov2015,peherstorferTransportbasedMultifidelityPreconditioner2019} struggle to achieve this correlation due to the inherent challenges of MCMC sampling. These existing methods often rely on sharing input samples between different fidelity levels, which is not straightforward in inverse problems where the model parameters from different fidelities originate from different posteriors. These limitations lead to suboptimal performance or even biased estimates. As detailed in our previous work~\cite{muchandimathSynchronizedStepMultilevel2025}, conventional coupling strategies struggle to maintain the correlation necessary for effective variance reduction in multi-fidelity estimators. To address these challenges, we present the following contributions in this paper:
\begin{enumerate}
    \item We develop a novel transport map-based coupling framework that transforms posterior distributions at different fidelities into a \textit{shared reference space}. This transformation enables us to efficiently generate correlated samples, overcoming the limitations of dissimilar and complex posterior distributions.
    \item We employ synchronized step correlation enhancement (SYNCE)~\cite{muchandimathSynchronizedStepMultilevel2025} as the coupling strategy in the reference space, a key departure from prior work that applied SYNCE directly in the parameter space. 
    \item The proposed methodology is modular and flexible, accommodating various transport map constructions and coupling strategies. This makes it broadly applicable to a wide range of multi-fidelity inverse problems.
    \item We validate our approach through numerical experiments on the SA turbulence model calibration for the challenging NACA~0012 airfoil test case at high angles of attack. We demonstrate $50\%$ reduction in computational cost while achieving accurate uncertainty quantification in model predictions.
\end{enumerate}
The remainder of the paper is organized as follows. In~\Cref{sec: methodology}, we introduce Bayesian inference and MCMC methods, multi-fidelity methods for variance reduction, and some background on couplings. In~\Cref{sec: proposed-method}, we present our novel transport map-based coupling framework and its implementation details. In~\Cref{sec: results}, we demonstrate our methodology on a simple toy problem and the NACA~0012 airfoil case. 

\section{Methodology}\label{sec: methodology}
In this section, we first introduce Bayesian inference and Markov chain Monte Carlo methods, which are used to sample from the posterior distribution of the model parameters to enable calibration. We then introduce multi-fidelity methods and show how they can be used to reduce the computational cost of MCMC. Finally, we discuss some coupling strategies for multi-fidelity MCMC, highlighting the challenges of achieving sample correlation between different fidelity levels in inverse problems.

\subsection{Bayesian inference and Markov chain Monte Carlo}\label{subsec: bayesian_inference_and_mcmc}
Bayesian inference treats model parameters $\bm{\theta} \in \mathbb{R}^d$ as random variables to be calibrated against experimental data $\bm{y_D}$, where $d$ is the dimension of the parameter set. The posterior distribution of the parameters $\pi(\bm{\theta})$ is given by Bayes' theorem as
\begin{align}\label{eq: bayes}
    \pi(\bm{\theta}) = \frac{\pi_{\text{likelihood}}(\bm{y_D}|\bm{\theta})\pi_{\text{prior}}(\bm{\theta})}{\pi_{\text{evidence}}(\bm{y_D})},
\end{align}
where $\pi_{\text{likelihood}}(\bm{y_D}|\bm{\theta})$ is the likelihood of the data given fixed parameters, $\pi_{\text{prior}}(\bm{\theta})$ is the prior distribution representing the state of knowledge about the parameters before observing the data, and $\pi_{\text{evidence}}(\bm{y_D})$ is the evidence term that normalizes the posterior distribution.

Assuming additive Gaussian observation noise $\bm{\epsilon} \sim \mathcal{N}(\bm{0},\bm{\Sigma})$, where $\bm{\Sigma}$ is the covariance matrix of the noise in the data and $\mathcal{N}$ represents a normal distribution, the likelihood function relates experimental data to model predictions according to $\bm{y}_D = f(x, \bm{\theta}) + \bm{\epsilon}$. The likelihood distribution is then given by
\begin{align}\label{eq: likelihood}
    \pi_{\text{likelihood}}(\bm{y_D}|\bm{\theta}) = \frac{1}{\sqrt{2\pi}^d\sqrt{\bm{\Sigma}}} \exp\left(-(\bm{y_D} - f(\bm{x}, \bm{\theta}))^T\bm{\Sigma}^{-1}(\bm{y_D} - f(\bm{x}, \bm{\theta}))\right),
\end{align}
where $d$ is the dimension of the parameter set. Generally, the posterior distribution $\pi(\bm{\theta})$ has no analytical form for non-linear models. 

MCMC methods~\cite{robertMonteCarloStatistical2004} generate samples from the posterior $\pi(\bm{\theta})$ by constructing a Markov chain with the posterior as its stationary distribution. The Metropolis--Hastings (MH) algorithm~\cite{hastingsMonteCarloSampling1970}, detailed in~\Cref{alg:mhmcmc}, is a widely used MCMC method\footnote{A variety of proposal mechanisms can be employed, ranging from simple random walks to sophisticated adaptive and gradient-based schemes, many of which are active areas of research~\cite{andrieuTutorialAdaptiveMCMC2008,cotter2013mcmc,robertslangevinDistributions1996}}. Each MCMC step requires evaluating the likelihood function -- and therefore the forward model -- making MCMC computationally prohibitive for expensive forward models.

\begin{algorithm}[H]
    \caption{Metropolis--Hastings MCMC algorithm~\cite{hastingsMonteCarloSampling1970}}~\label{alg:mhmcmc}
    \begin{algorithmic}[1]
        \STATE \textbf{Input:} 
        \STATE \hspace{1em} Target distribution $\pi(\bm{\theta})$
        \STATE \hspace{1em} Proposal distribution $q(\cdot \mid \cdot)$
        \STATE \hspace{1em} Initial guess $\bm{\theta}^0$
        \STATE \hspace{1em} Number of samples $N$
        \vspace{1em}
        \STATE \textbf{Output:} 
        \STATE \hspace{1em} Samples $\{\boldsymbol{\theta}^i\}_{i=1}^{N}$
        \vspace{1em}
        \FOR{$i = 0$ to $N-1$}
            \STATE \textbf{Step 1: Generate proposal}
            \STATE Sample candidate $\bm{\theta}^* \sim q(\cdot \mid \bm{\theta}^i)$ \hfill \textit{(proposal step)}
            \vspace{1em}
            \STATE \textbf{Step 2: Accept/reject proposal}
            \STATE Sample $u \sim \mathcal{U}[0,1]$
            \STATE Compute acceptance ratio:
                \begin{align*}
                    \alpha = \min\left(1, \frac{\pi\left(\bm{\theta}^*\right)q\left(\bm{\theta}^i \mid \bm{\theta}^*\right)}{\pi\left(\bm{\theta}^i\right)q\left(\bm{\theta}^* \mid \bm{\theta}^i\right)}\right)
                \end{align*}
                \STATE Set $\bm{\theta}^{i+1} = \begin{cases}
                    \bm{\theta}^* & \text{with probability } \alpha \\
                    \bm{\theta}^i & \text{with probability } 1-\alpha
                    \end{cases}$ \hfill \textit{(accept/reject step)}
        \ENDFOR
        \vspace{1em}
        \RETURN $\{\boldsymbol{\theta}^i\}_{i=1}^{N}$
    \end{algorithmic}
\end{algorithm}

Posterior samples generated by MCMC are used to quantify uncertainty in the model predictions. For an output functional of interest $Q(\bm{\theta})$, such as lift or drag coefficient, the posterior distribution of the quantity of interest (QOI) is estimated by propagating the posterior samples through the output functional. It is important to note that the QOI may differ from the model outputs used for calibration. This distinction highlights the need for comprehensive characterization of parameter uncertainty, so it can be effectively propagated to various outputs of interest beyond the calibration data itself. The posterior mean of the QOI is then estimated using Monte Carlo (MC) integration as
\begin{align}\label{eq: mc-mean}
    \hat{Q}^{MC} = \frac{1}{N}\sum_{i=1}^{N}Q(\bm{\theta}^i), \quad \bm{\theta}^i \sim \pi(\bm{\theta}),
\end{align}
where $\bm{\theta}^i \sim \pi(\bm{\theta})$ denotes posterior samples and $N$ denotes the number of samples used in the estimation process. The MC estimator is unbiased, and the variance of the estimator is given by 
\begin{align}\label{eq: mc-variance}
    \mathbb{V}\text{ar}\left[\hat{Q}^{MC}\right] = \frac{1}{N}\mathbb{V}\text{ar}\left[Q\right].
\end{align}
The fundamental challenge is that accurate estimation requires large sample sizes, but each posterior sample involves an expensive model evaluation through the likelihood. Surrogate models are often employed both in the calibration stage and in the subsequent MC propagation stage to alleviate this computational burden. However, surrogate approximations introduce biases in both stages --- distorting the inferred posterior as well as the estimated uncertainty in the QOI, leading to multiple sources of error~\cite{rayBayesianParameterEstimation2016,daronchSensitivityCalibrationTurbulence2020}.

\subsection{Multi-fidelity methods for variance reduction}\label{subsec: multi-fidelity_methods}

Multi-fidelity techniques strategically combine an ensemble of computational models with different accuracy and cost to achieve better estimates than using a single model alone. The ensemble consists of a high-fidelity model that is expensive to evaluate and low-fidelity models that are cheaper to evaluate but less accurate. The low-fidelity model can be a surrogate, reduced order model, or a coarse grid solution that approximates the input-output relationship of the high-fidelity model. A plethora of multi-fidelity methods have been proposed in the literature~\cite{peherstorferSurveyMultifidelityMethods2018,nelson1990control,gorodetskyGeneralizedApproximateControl2020,thomasgroupedACV2024,gilesMultilevelMonteCarlo2015,schaden2020multilevel}. The most common approach is to use a linear weighted combination of the high-fidelity and low-fidelity models to obtain a more accurate estimate of the quantity of interest. 

In this paper, we consider a hierarchy of models of increasing fidelity $f_{\ell} = f(\bm{x}_{\ell}, \bm{\theta}_{\ell})$, where $\ell=0,1,\ldots,L-1$ for the low-fidelity models and $\ell=L$ for the high-fidelity model. Each model $f_{\ell}$ corresponds to a different likelihood and a different posterior distribution $\pi_{\ell} = \pi(\bm{\theta}_{\ell})$ when calibrated against the same data. Similarly, we refer to $Q_{\ell}(\bm{\theta}_{\ell})$ as the quantity of interest for the $\ell$'th fidelity model. 

Control variates (CV)~\cite{nelson1990control}, approximate control variates (ACV)~\cite{gorodetskyGeneralizedApproximateControl2020} and multi-level Monte Carlo (MLMC)~\cite{gilesMultilevelMonteCarlo2015} are popular multi-fidelity methods that focus on variance reduction. MLMC estimators exploit a telescoping sum formulation and can be derived from the ACV framework by setting each control variate weight to be negative one. The MLMC estimator for the posterior mean of the QOI is given by
\begin{align}\label{eq: mlmc-esimate}
    \hat{Q}^{\text{MLMC}}_L &= \hat{Q}_{0}+ \sum_{\ell=1}^{L} \left(\hat{Q}_{\ell} - \hat{Q}_{\ell-1}\right),
\end{align}
where $\hat{Q}_{\ell}$ is the MC estimator for the $\ell$'th fidelity model. The key to variance reduction in multi-fidelity methods is high correlation between the low-fidelity and high-fidelity outputs~\cite{gorodetskyGeneralizedApproximateControl2020}. Consider a simple bi-fidelity example by setting $L=1$, then
\begin{align}\label{eq: mlmc-bifidelity}
    \hat{Q}^{\text{MLMC}}_1 &= \hat{Q}_{0} + \left(\hat{Q}_{1} - \hat{Q}_{0}\right), \\ \nonumber
    &= \frac{1}{N_{0}}\sum_{i=1}^{N_{0}}Q_{0}(\bm{\theta}_{0}^i) + \frac{1}{N_1}\sum_{i=1}^{N_{1}}\left(Q_{1}(\bm{\theta}_{1}^i) - Q_{0}(\bm{\vartheta}_{0}^i)\right),
\end{align}
where $\bm{\theta}^i_0$ denotes samples from the low-fidelity posterior $\pi_0$, $\bm{\theta}^i_1$ denotes samples from the high-fidelity posterior $\pi_1$, and $\bm{\vartheta}^i_0$ denotes samples from the low-fidelity posterior $\pi_0$ that maybe different from the samples $\bm{\theta}^i_0$ used in the first term. For simplicity, we assume $\mathbb{V}\text{ar}\left[Q_1(\bm{\theta}_1)\right] \approx \mathbb{V}\text{ar}\left[Q_0(\bm{\vartheta}_0)\right] = V$, where $\bm{\theta}_1 = \left\{\bm{\theta}^i_1\right\}_{i\in \mathbb{N}}$ and $\bm{\vartheta}_0 = \left\{\bm{\vartheta}^i_0\right\}_{i\in \mathbb{N}}$, $\mathbb{N} = \left\{1,2,3,\ldots\right\}$. We can then express the variance of the MLMC estimator as
\begin{align}\label{eq: mlmc-variance}
    \mathbb{V}\text{ar}&\left[\hat{Q}^{\text{MLMC}}_1\right] = \frac{1}{N_0}V + \frac{1}{N_1}2V(1-\rho), \\ \nonumber
    \rho &= \mathbb{C}\text{ov}\left[Q_1(\bm{\theta}_{1}), Q_0(\bm{\vartheta}_{0})\right] / \sqrt{\mathbb{V}\text{ar}\left[Q_1(\bm{\theta}_{1})\right] \mathbb{V}\text{ar}\left[Q_0(\bm{\vartheta}_{0})\right]},
\end{align}
where $\rho$ is the Pearson correlation coefficient between the low-fidelity and high-fidelity outputs. The correlation coefficient $\rho$ directly controls the estimator efficiency through the coupled variance term. When $\rho\approx1$, the coupled variance term in the RHS approaches zero, enabling dramatic variance reduction. Consequently, for a fixed target variance, the number of expensive high-fidelity samples $N_1$ decreases substantially, while maintaining $N_0 \gg N_1$ through optimal sample allocation~\cite{gilesMultilevelMonteCarlo2015}. To reiterate, the key to achieving variance reduction is to have high correlation between the low-fidelity and high-fidelity outputs~\cite{gorodetskyGeneralizedApproximateControl2020}.

When we have more than two-fidelity levels, the MLMC framework extends naturally through the telescoping sum formulation in~\Cref{eq: mlmc-esimate}. Each difference term $\hat{Q}_{\ell} - \hat{Q}_{\ell-1}$ in the sum requires correlated samples from the $\ell$'th and $\ell-1$'th fidelity levels. We do not need to correlate samples across all levels simultaneously --- only between adjacent levels. This pairwise correlation requirement simplifies the coupling problem significantly by focusing only on the $\ell$ and $\ell-1$ fidelity levels while still enabling effective variance reduction. Crucially, since each difference term is statistically independent, these computations can be performed in parallel, further enhancing computational efficiency.

In forward problems, correlation between adjacent levels is easily achieved by sharing input samples of the high-fidelity model with all the low-fidelity models. Inverse problems are more challenging. The model parameters are unknown a priori, and MCMC methods introduce dependencies on the forward model. This complicates direct sample sharing~\cite{dodwellHierarchicalMultilevelMarkov2015,muchandimathSynchronizedStepMultilevel2025}, and, achieving correlation between adjacent fidelity level outputs becomes difficult. To ensure unbiased multilevel estimators, samples at each fidelity level must be drawn from their corresponding posterior distributions $\bm{\theta}_{\ell} \sim \pi_{\ell}$. The challenge is to create \textit{correlated samples between adjacent fidelities} that are consistent with the posterior distributions of the respective models. We address this challenge by using \textit{coupling} techniques that we discuss next.

\subsection{Couplings}\label{subsec: couplings}
A coupling~\cite{villaniOptimalTransportOld} between two probability measures $\mu$ and $\nu$ on the measurable spaces $(X, \mathcal{X})$ and $(Y, \mathcal{Y})$ is a joint probability measure $\gamma$ on the product space $(X \times Y, \mathcal{X} \otimes \mathcal{Y})$ such that the marginal distributions of $\gamma$ are $\mu$ and $\nu$, i.e., $\gamma(A \times Y) = \mu(A)$ for all $A \in \mathcal{X}$ and $\gamma(X \times B) = \nu(B)$ for all $B \in \mathcal{Y}$. There are broadly two classes of couplings: deterministic couplings and non-deterministic couplings~\cite{villaniOptimalTransportOld}.

\textit{Non-deterministic couplings} use probabilistic mechanisms to couple marginal distributions to create a joint distribution. These couplings connect the marginal distributions through shared randomness or correlated sampling techniques. Popular examples are maximal couplings~\cite{Thorissonmaximal1986} that maximize the probability of coincidence between samples from the two marginals, reflection couplings~\cite{jacob2019unbiasedmarkovchainmonte} that use reflections based on shared randomness to create correlation, and, common random number (CRN) couplings~\cite{jacob2019unbiasedmarkovchainmonte} that use common random numbers in sampling methods.

\textit{Deterministic couplings} are characterized by measurable deterministic functions between samples from two distributions. Formally, they are described through a measurable map $T:X \rightarrow Y$ such that $Y=T(X)$ and $\nu = T_\#\mu$, where $T_\#$ is termed as the push-forward of $\mu$ by $T$~\cite{villaniOptimalTransportOld}. The deterministic function $T$ is called a transport map, and the coupling takes the form $\gamma=\left(\text{Id},T\right)_{\#}\mu$, where $\text{Id}$ is the identity mapping. Popular examples of deterministic couplings are optimal transport maps~\cite{villaniOptimalTransportOld,peyreComputationalOptimalTransport2019} and Knothe--Rosenblatt rearrangements~\cite{knottOptimalMappingDistributions1984}.

\subsubsection{Couplings in MCMC sampling}\label{subsubsec: coupled_mcmc_sampling}
Both types of couplings have been explored in the context of MCMC sampling for inverse problems but address distinct objectives. Non-deterministic couplings have been used to reduce bias in MCMC estimates~\cite{jacob2019unbiasedmarkovchainmonte} and to improve correlation in multi-fidelity MCMC estimators~\cite{muchandimathSynchronizedStepMultilevel2025}. Deterministic couplings have been used primarily to improve MCMC sampling efficiency by transforming the target distribution into a simpler reference distribution~\cite{marzoukSamplingMeasureTransport2016,parnoTransportMapAccelerated2018}.

\textit{Non-deterministic couplings in MCMC sampling:}
MCMC samples are coupled by coupling the Markov kernels that generate the samples.~\citet{olearyMetropolisHastingsTransitionKernel2023} define a coupling of two ergodic Markov kernels $\mathcal{K}_X: X \times \mathcal{X} \rightarrow [0, 1]$ with measure $\mu$, and $\mathcal{K}_Y: Y \times \mathcal{Y} \rightarrow [0, 1]$ with measure $\nu$, as a joint Markov kernel $\mathcal{K}: (X \times Y) \times (\mathcal{X} \otimes \mathcal{Y}) \rightarrow [0, 1]$ with joint measure $\gamma$ satisfying the following property:
\begin{align*}
  \int_{Y}^{}\mathcal{K}((x,y),A \times dy) &= \mathcal{K}_X(x,A),\\
  \int_{X}^{}\mathcal{K}((x,y),dx \times B) &= \mathcal{K}_Y(y,B).
\end{align*}
The above marginal property ensures that the coupled Markov kernel $\mathcal{K}$ preserves the marginal distributions $\mu$ and $\nu$ of the individual kernels $\mathcal{K}_X$ and $\mathcal{K}_Y$. 

Translating this abstract framework into practical algorithms requires coupling both the proposal and acceptance mechanisms~\cite{olearyMetropolisHastingsTransitionKernel2023}. The key insight is that MCMC kernel coupling essentially requires two components: generating correlated proposal points through a joint proposal distribution, and using a common uniform random number to accept/reject the proposals to maintain the Markov property. Therefore, to achieve high correlation between different fidelity levels, we need to design effective coupling strategies for the proposal distributions that generate correlated proposals. Rather than sampling $\bm{\theta}^*_{\ell} \sim q_{\ell}(\cdot \mid \bm{\theta}_{\ell})$ and $\bm{\theta}^*_{\ell-1} \sim q_{\ell-1}(\cdot \mid \bm{\theta}_{\ell-1})$ independently, we sample the pair $(\bm{\theta}_{\ell}^*,\bm{\theta}^*_{\ell-1})$ jointly from a coupled proposal distribution $\gamma_{\ell}\left((\cdot, \cdot) \mid (\bm{\theta}_{\ell},\bm{\theta}_{\ell-1})\right)$ whose marginals are the proposal distributions $q_{\ell} = \int \gamma_{\ell} d\bm{\theta}_{\ell-1}$ and $q_{\ell-1} = \int \gamma_{\ell} d\bm{\theta}_{\ell}$. Algorithm~\ref{alg:coupling-mlmc} outlines a typical coupled MCMC algorithm for a plug-in coupling strategy $\gamma_{\ell}$ between the $\ell$'th and $\ell-1$'th fidelity levels. 
\begin{algorithm}[ht]
    \caption{Coupled MCMC algorithm~\cite{muchandimathSynchronizedStepMultilevel2025}}
    \label{alg:coupling-mlmc}
    \begin{algorithmic}[1]
        \STATE \textbf{Input:} 
        \STATE \hspace{1em} Target distributions $\pi_{\ell}(\bm{\theta}_{\ell}), \pi_{\ell-1}(\bm{\theta}_{\ell-1})$
        \STATE \hspace{1em} Coupled proposal distribution $\gamma_{\ell}((\cdot, \cdot) \mid (\cdot, \cdot))$
        \STATE \hspace{1em} Initial guess $\bm{\theta}_{\ell}^0,\bm{\theta}_{\ell-1}^0$
        \STATE \hspace{1em} Number of samples $N$
        \vspace{1em}
        \STATE \textbf{Output:} 
        \STATE \hspace{1em} Coupled correlated samples $\{\boldsymbol{\theta}_{\ell}^i, \boldsymbol{\theta}_{\ell-1}^i\}_{i=1}^{N}$
        \vspace{1em}
        \FOR{$i = 0$ to $N-1$}
            \STATE \textbf{Step 1: Generate coupled proposals}
            \STATE Sample coupled candidate $\left(\bm{\theta}_{\ell}^*, \bm{\theta}_{\ell-1}^*\right) \sim \gamma_{\ell}\left((\cdot, \cdot) \mid (\bm{\theta}_{\ell}^i, \bm{\theta}_{\ell-1}^i)\right)$ \hfill \textit{(proposal coupling step)}
            \vspace{1em}
            \STATE \textbf{Step 2: Accept/reject proposals marginally}
            \STATE Sample common uniform random number $u \sim \mathcal{U}[0,1]$ \hfill \textit{(acceptance ratio coupling step)}
            \FOR{$j=\ell; \ell-1$}
                \STATE Compute acceptance ratio for both levels:
                \begin{align*}
                    \alpha_{j} &= \min\left(1, \frac{\pi_{j}\left(\bm{\theta}_{j}^*\right) q_{j}\left(\bm{\theta}_{j}^i \mid \bm{\theta}_{j}^*\right)}{\pi_{j}\left(\bm{\theta}_{j}^i\right)q_{j}\left(\bm{\theta}_{j}^* \mid \bm{\theta}_{j}^i\right)}\right) \\ 
                \end{align*} 
                \STATE Set $\bm{\theta}_{j}^{i+1} = \begin{cases}
                    \bm{\theta}_j^* & \text{with probability } \alpha_j \\
                    \bm{\theta}_j^i & \text{with probability } 1-\alpha_j
                    \end{cases}$ \hfill \textit{(marginal accept/reject step)}
            \ENDFOR
        \ENDFOR
        \vspace{1em}
        \RETURN $\{\boldsymbol{\theta}_{\ell}^i, \boldsymbol{\theta}_{\ell-1}^i\}_{i=1}^{N}$
    \end{algorithmic}
\end{algorithm}
The effectiveness of coupled MCMC algorithms depends critically on the design of the coupled proposal distribution $\gamma_{\ell} \left((\cdot, \cdot) \mid (\bm{\theta}_{\ell},\bm{\theta}_{\ell-1})\right)$. 

Several non-deterministic couplings strategies have been developed and applied for this purpose. Traditional couplings (maximal~\cite{Thorissonmaximal1986}, reflection~\cite{jacob2019unbiasedmarkovchainmonte}, delayed acceptance~\cite{dodwellHierarchicalMultilevelMarkov2015}) have been successfully applied to multi-fidelity MCMC sampling by creating joint proposal distributions that enforce identical or highly correlated proposals across fidelity levels. Existing methods, though different in their implementation, typically force the same or highly correlated proposals for both fidelity levels leading to high correlation only when the posterior distributions of the two fidelity levels are similar. However, in practice, the posterior distributions of different fidelity levels can be different, leading to low acceptance rates and poor correlation. We proposed the synchronized step correlation enhancement (SYNCE) coupling strategy to overcome the limitation of poor correlation in dissimilar posteriors by using common random number (CRN) couplings in our previous work~\cite{muchandimathSynchronizedStepMultilevel2025}. The method is simple and works by drawing a shared random variable $\bm{\eta} \sim \mathcal{N}(\bm{0}, \bm{C})$ and uses this shared random number to propose the same direction and magnitude of change for both fidelity levels. The covariance $\bm{C}$ controls the step size and the correlation structure. For Gaussian random walk proposals, the SYNCE coupling takes the form
\begin{align}\label{eq: synce-rw-coupling}
    \bm{\theta}^*_{\ell} = \bm{\theta}^i_{\ell} + \bm{\eta}, \quad \bm{\theta}^*_{\ell-1} = \bm{\theta}^i_{\ell-1} + \bm{\eta},
\end{align}
and represents the proposal coupling step (Step 1) in~\Cref{alg:coupling-mlmc}. We recommend~\cite{muchandimathSynchronizedStepMultilevel2025} for a detailed description of the SYNCE coupling strategy and its theoretical properties.

\begin{figure}[ht]
    \centering
    \def\svgwidth{0.7\textwidth}
    %% Creator: Inkscape 1.4.2 (ebf0e940, 2025-05-08), www.inkscape.org
%% Accompanies image file 'transport_map.pdf' (pdf, eps, ps)
\begingroup%
  \makeatletter%
  \providecommand\color[2][]{%
    \errmessage{(Inkscape) Color is used for the text in Inkscape, but the package 'color.sty' is not loaded}%
    \renewcommand\color[2][]{}%
  }%
  \providecommand\transparent[1]{%
    \errmessage{(Inkscape) Transparency is used (non-zero) for the text in Inkscape, but the package 'transparent.sty' is not loaded}%
    \renewcommand\transparent[1]{}%
  }%
  \providecommand\rotatebox[2]{#2}%
  \newcommand*\fsize{\dimexpr\f@size pt\relax}%
  \newcommand*\lineheight[1]{\fontsize{\fsize}{#1\fsize}\selectfont}%
  \ifx\svgwidth\undefined%
    \setlength{\unitlength}{708.24102783bp}%
    \ifx\svgscale\undefined%
      \relax%
    \else%
      \setlength{\unitlength}{\unitlength * \real{\svgscale}}%
    \fi%
  \else%
    \setlength{\unitlength}{\svgwidth}%
  \fi%
  \global\let\svgwidth\undefined%
  \global\let\svgscale\undefined%
  \makeatother%
  \begin{picture}(1,0.71437826)%
    \lineheight{1}%
    \setlength\tabcolsep{0pt}%
    \put(0,0){\includegraphics[width=\unitlength,page=1]{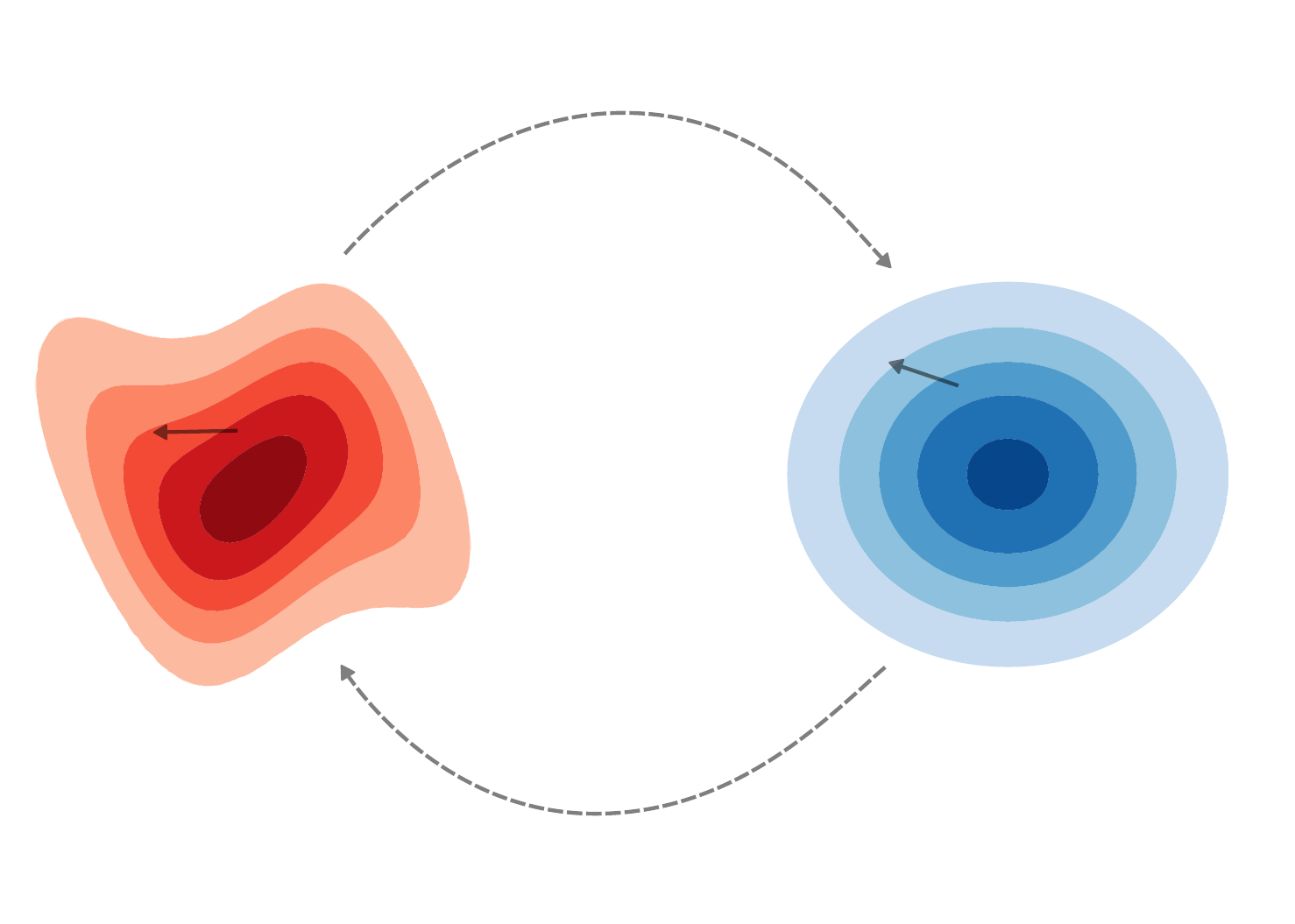}}%
    \put(0.46161929,0.00860902){\color[rgb]{0,0,0}\makebox(0,0)[lt]{\lineheight{1.25}\smash{\begin{tabular}[t]{l}$T^{-1}(\cdot)$\end{tabular}}}}%
    \put(0.45073966,0.67340938){\color[rgb]{0,0,0}\makebox(0,0)[lt]{\lineheight{1.25}\smash{\begin{tabular}[t]{l}$T(\cdot)$\end{tabular}}}}%
    \put(0.14941375,0.12676934){\color[rgb]{0,0,0}\makebox(0,0)[lt]{\lineheight{1.25}\smash{\begin{tabular}[t]{l}$\pi$\end{tabular}}}}%
    \put(0.76040198,0.13971474){\color[rgb]{0,0,0}\makebox(0,0)[lt]{\lineheight{1.25}\smash{\begin{tabular}[t]{l}$\pi_r$\end{tabular}}}}%
    \put(0,0){\includegraphics[width=\unitlength,page=2]{figures/transport_map.pdf}}%
  \end{picture}%
\endgroup%

    \caption{Schematic of transport map enhanced MCMC sampling. The complex target distribution $\pi$ is mapped to a simpler reference distribution $\pi_r$ using the transport map $T$. Sampling is performed in the reference space using a simple proposal distribution, and samples are mapped back to the target space using the inverse transport map $T^{-1}$~\cite{parnoTransportMapAccelerated2018}.}
    \label{fig: transportmap}
\end{figure}

\textit{Deterministic couplings in MCMC sampling:}
In contrast to non-deterministic approaches that link different posterior distributions, deterministic approaches in MCMC couple a complex posterior distribution $\pi$ to a simpler reference distribution $\pi_r$ (typically a Gaussian)~\cite{parnoTransportMapAccelerated2018}. Sampling is then performed in the Gaussianized reference space that is easier to explore using a reference space proposal. From another perspective, transport maps convert simple reference space proposals into non-Gaussian proposals that adapt to the target posterior's geometry. A schematic illustrating the transport map approach is shown in~\Cref{fig: transportmap}. The transport map $T$ is constructed to approximate the pushforward of the target measure to the reference measure $\mu_r \approx T_{\#}\mu$, where $\mu_r$ is the reference measure and $\mu$ is the target measure\footnote{We assume that both the target and reference measures are absolutely continuous on $\mathbb{R}^d$, with densities $\pi$ and $\pi_r$ respectively}. An algorithmic outline of a transport map enhanced MCMC method is shown in~\Cref{alg:tm-mcmc}. Crucially, the proposal distribution in the target space $q_{\theta}\left(\bm{\theta}^* \mid \bm{\theta}^i\right) = q_{r}\left(T\left(\bm{\theta}^*\right) \mid T\left(\bm{\theta}^i\right)\right)\left| \det J_T \left(\bm{\theta}^*\right) \right|$ is given by the change of variables formula, where $J_T$ is the Jacobian of the transport map $T$.

These methods have proven highly effective for single-fidelity problems~\cite{marzoukSamplingMeasureTransport2016}, with transport map MCMC achieving orders of magnitude improvements in sampling efficiency over traditional MCMC methods~\cite{parnoTransportMapAccelerated2018,gabrieAdaptiveMonteCarlo2022,greniouxSamplingApproximateTransport2024}. The success stems from the transport map's ability to adapt to complex posterior geometries, effectively creating geometrically aware proposals. However, the application of transport maps to multi-fidelity problems remains relatively unexplored. Only~\citet{peherstorferTransportbasedMultifidelityPreconditioner2019} have focused on using transport maps in a multi-fidelity context, where they use low-fidelity models to construct transport maps that precondition MCMC sampling in the high-fidelity model. While this approach improves sampling efficiency for the high-fidelity model -- provided that the low-fidelity model is a good approximation of the high-fidelity model -- it does not directly create correlation between different fidelity levels, which is essential for effective variance reduction in multi-fidelity estimators. Moreover, in practice, the low-fidelity posterior often serves as a poor preconditioner even when the low-fidelity model itself provides a reasonable physical approximation --- leading to degraded performance characterized by low acceptance rates and inefficient exploration.

\begin{algorithm}[H]
    \caption{MCMC with transport map enhanced proposals~\cite{parnoTransportMapAccelerated2018}}
    \label{alg:tm-mcmc}
    \begin{algorithmic}[1]
        \STATE \textbf{Input:} 
        \STATE \hspace{1em} Target distribution $\pi(\bm{\theta})$
        \STATE \hspace{1em} Reference space proposal distribution $q_r(\cdot \mid \cdot)$
        \STATE \hspace{1em} Transport map $T$ 
        \STATE \hspace{1em} Initial guess $\bm{\theta}^0$
        \STATE \hspace{1em} Number of samples $N$
        \vspace{1em}
        \STATE \textbf{Output:} 
        \STATE \hspace{1em} Samples $\{\boldsymbol{\theta}^i\}_{i=1}^{N}$
        \vspace{1em}
        \FOR{$i = 0$ to $N-1$}
            \STATE \textbf{Step 1: Forward map to reference space}
            \STATE Compute $\bm{r}^i = T(\bm{\theta}^i)$ \hfill \textit{(forward map)}
            \vspace{1em}
            \STATE \textbf{Step 2: Generate proposal in reference space}
            \STATE Sample candidate $\bm{r}^* \sim q_r(\cdot \mid \bm{r}^i)$ \hfill \textit{(proposal step)}
            \vspace{1em}
            \STATE \textbf{Step 3: Inverse map to target space}
            \STATE Compute $\bm{\theta}^* = T^{-1}(\bm{r}^*)$ \hfill \textit{(inverse map)}
            \vspace{1em}
            \STATE \textbf{Step 4: Accept/reject proposal}
            \STATE Sample $u \sim \mathcal{U}[0,1]$
            \STATE Compute acceptance ratio: \hfill \textit{(Jacobian adjustment)}
                \begin{align*}
                    \alpha = \min\left(1, \frac{\pi\left(\bm{\theta}^*\right)q_r\left(T\left(\bm{\theta}^i\right) \mid T\left(\bm{\theta}^*\right)\right)\left|\det J_T\left(\bm{\theta}^i\right)\right|}{\pi\left(\bm{\theta}^i\right)q_r\left(T\left(\bm{\theta}^*\right) \mid T\left(\bm{\theta}^i\right)\right)\left|\det J_T\left(\bm{\theta}^*\right)\right|}\right)
                \end{align*} 
                \STATE Set $\bm{\theta}^{i+1} = \begin{cases}
                    \bm{\theta}^* & \text{with probability } \alpha \\
                    \bm{\theta}^i & \text{with probability } 1-\alpha
                    \end{cases}$ \hfill \textit{(accept/reject step)}
        \ENDFOR
        \vspace{1em}
        \RETURN $\{\boldsymbol{\theta}^i\}_{i=1}^{N}$
    \end{algorithmic}
\end{algorithm}

This separation between geometric adaptation and cross-fidelity correlation represents a significant missed opportunity. Transport maps possess the geometric adaptability needed to handle complex posterior distributions, while multi-fidelity methods require correlation structures between fidelity levels to achieve variance reduction. Existing literature handles these two aspects separately, with neither approach fully addressing the challenge of multi-fidelity MCMC sampling --- adapting to complex geometries while ensuring high correlation between fidelity levels. In the following section, we present our novel approach that addresses this gap by combining the geometric advantages of transport maps with correlation-enhancing strategies like SYNCE to create effective multi-fidelity couplings.

\section{Proposed Method}\label{sec: proposed-method}

We propose a novel approach that combines the strengths of deterministic and non-deterministic couplings through a \textit{shared reference space}. The core idea lies in using deterministic transport maps to transform complex posterior distributions from successive fidelity levels into a shared simplified reference space. Once in this shared space, we apply a non-deterministic coupling method to generate correlated proposals before mapping them back into the original target spaces. By combining these two components, our method addresses the limitations of existing approaches: non-deterministic couplings often perform poorly in complex target spaces with disparate geometries~\cite{dodwellHierarchicalMultilevelMarkov2015,muchandimathSynchronizedStepMultilevel2025}, while deterministic transport approaches focus on single-fidelity efficiency without enhancing multi-fidelity correlation~\cite{peherstorferTransportbasedMultifidelityPreconditioner2019}. Our framework leverages transport-based simplification to create a favorable environment where correlation-enhancing couplings can operate more effectively.

\begin{algorithm}[h!]
    \caption{TM-SYNCE coupling}\label{alg:tm-synce-coupling}
    \begin{algorithmic}[1]
        \STATE \textbf{Input:} 
        \STATE \hspace{1em} Current samples $(\bm{\theta}_{\ell}^i, \bm{\theta}_{\ell-1}^i)$
        \STATE \hspace{1em} Transport maps $(T_{\ell}, T_{\ell-1})$
        \STATE \hspace{1em} Reference covariance $\bm{C}_r$
        \STATE \hspace{1em} Resynchronization parameter $\omega_{\ell} \in [0,1]$
        \vspace{1em}
        \STATE \textbf{Output:} 
        \STATE \hspace{1em} Coupled proposals $(\bm{\theta}_{\ell}^*, \bm{\theta}_{\ell-1}^*)$
        \vspace{1em}
        \STATE \textbf{Step 1: Forward mapping to reference space}
        \STATE $\bm{r}_{\ell}^i = \begin{cases}
            T_{\ell}(\bm{\theta}_{\ell}^i) & \text{if direct} \\
            (T_{\ell-1} \circ T_{\ell})(\bm{\theta}_{\ell}^i) & \text{if deep}
            \end{cases}$
            \STATE $\bm{r}_{\ell-1}^i = T_{\ell-1}(\bm{\theta}_{\ell-1}^i)$
        \vspace{1em}
        \STATE \textbf{Step 2: Generate coupled proposals in reference space}
        \STATE Sample $\bm{\eta}_r \sim \mathcal{N}(\bm{0}, \bm{C}_r)$
        \STATE $(\bm{r}_{\ell}^*, \bm{r}_{\ell-1}^*) = \begin{cases}
        (\bm{\eta}_r, \bm{\eta}_r) & \text{with probability } \omega_{\ell} \text{ (independent)} \\
        (\bm{r}_{\ell}^i + \bm{\eta}_r, \bm{r}_{\ell-1}^i + \bm{\eta}_r) & \text{with probability } 1-\omega_{\ell} \text{ (random walk)}
        \end{cases}$
        \vspace{1em}
        \STATE \textbf{Step 3: Inverse mapping to target spaces}
        \STATE $\bm{\theta}_{\ell}^* = \begin{cases}
            T_{\ell}^{-1}(\bm{r}_{\ell}^*) & \text{if direct} \\
            (T_{\ell-1} \circ T_{\ell})^{-1}(\bm{r}_{\ell}^*) & \text{if deep}
            \end{cases}$
            \STATE $\bm{\theta}_{\ell-1}^* = T_{\ell-1}(\bm{r}_{\ell-1}^*)$
        \vspace{1em}
        \RETURN $(\bm{\theta}_{\ell}^*, \bm{\theta}_{\ell-1}^*)$
    \end{algorithmic}
\end{algorithm}

~\Cref{alg:tm-synce-coupling} provides the complete implementation of the transport map-SYNCE (TM-SYNCE) coupling strategy that combines transport maps with the SYNCE coupling mechanism and represents the proposal coupling step (Step 1) in~\Cref{alg:coupling-mlmc}. Given transport maps $T_{\ell}$ and $T_{\ell-1}$ for the $\ell$'th and $\ell-1$'th fidelity levels respectively, we implement the following steps at each iteration of the coupled MCMC algorithm:
\begin{enumerate}
    \item Forward map to reference space: Map the current samples from both fidelity levels to a common reference space using their respective maps
    \begin{align}
        \bm{r}_\ell^i &= T_\ell(\bm{\theta}_\ell^i), \quad 
        \bm{r}_{\ell-1}^i = T_{\ell-1}(\bm{\theta}_{\ell-1}^i).
    \end{align}
    \item Coupled proposal in reference space: Generate correlated proposals in the reference space using a non-deterministic coupling strategy
    \begin{align}
        \left(\bm{r}_{\ell}^*, \bm{r}_{\ell-1}^*\right) \sim \gamma_{r}\left((\cdot, \cdot) \mid (\bm{r}_{\ell}^i, \bm{r}_{\ell-1}^i)\right).
    \end{align}
    \item Inverse map to target spaces: Map the correlated proposals back to the target spaces using the inverse of their respective transport maps
    \begin{align}
        \bm{\theta}_\ell^i &= T^{-1}_\ell(\bm{r}_\ell^i), \quad 
        \bm{\theta}_{\ell-1}^i = T^{-1}_{\ell-1}(\bm{r}_{\ell-1}^i).
    \end{align}
\end{enumerate}
After obtaining the coupled proposals $(\bm{\theta}_\ell^*, \bm{\theta}_{\ell-1}^*)$, we perform standard marginal Metropolis-Hastings acceptance at each fidelity using a shared uniform random number (Step 2 in~\Cref{alg:coupling-mlmc}). The subsequent subsections discuss in detail each component referenced in~\Cref{alg:tm-synce-coupling}.
\subsection{Reference-space coupling via SYNCE}\label{subsec: reference-synce}
Within the shared reference space, we denote by $\gamma_r((\cdot,\cdot)\mid(\bm r_\ell^i,\bm r_{\ell-1}^i))$ a generic coupling distribution that proposes a correlated pair $(\bm r_\ell^*,\bm r_{\ell-1}^*)$ given current states $(\bm r_\ell^i,\bm r_{\ell-1}^i)$. In this work, we instantiate $\gamma_r$ with the SYNCE strategy, which uses common random number couplings to strengthen inter-fidelity correlation. Importantly, operating on the simplified reference geometry allows SYNCE to overcome difficulties it faces when applied directly in the target space to high-dimensional or multi-modal target distributions.

Concretely, SYNCE in the reference space alternates between two proposal mechanisms that represent the classic exploration-exploitation trade-off in MCMC sampling. The first mechanism is a random walk proposal (\Cref{eq: synce-rw-coupling}) that is centered around the current sample, promoting exploration of the posterior structure. The second mechanism is an independent proposal that is not dependent on the current sample, allowing for exploitation of high density regions that are well captured by the transport maps and help resynchronize chains when they diverge. The two mechanisms are controlled by a resynchronization parameter $\omega\in[0,1]$ that sets the frequency of independent steps~\cite{muchandimathSynchronizedStepMultilevel2025}. The SYNCE coupling in the reference space is given by:
\begin{align}
    \text{(Independent / exploit)}\quad
    &\bm\eta_r \sim \mathcal N(\bm 0,\bm C_r),\quad 
      \bm r_\ell^*=\bm\eta_r,\ \ \bm r_{\ell-1}^*=\bm\eta_r \quad \text{with prob. } \omega, \label{eq: synce-ind} \\
    \text{(Random-walk / explore)}\quad
    &\bm\eta_r \sim \mathcal N(\bm 0,\bm C_r),\quad 
      \bm r_\ell^*=\bm r_\ell^i+\bm\eta_r,\ \ \bm r_{\ell-1}^*=\bm r_{\ell-1}^i+\bm\eta_r \quad \text{with prob. } 1-\omega, \label{eq: synce-rw}
\end{align}
where $\bm C_r$ is a reference covariance. The shared noise $\bm\eta_r$ induces strong positive correlation across fidelities, while the independent steps prevent desynchronization when posteriors differ substantially~\cite{muchandimathSynchronizedStepMultilevel2025}. After proposing $(\bm r_\ell^*,\bm r_{\ell-1}^*)$, we map back to the respective target spaces via the inverse maps and perform standard marginal Metropolis-Hastings acceptance at each fidelity using a shared uniform random number (Step 2 in~\Cref{alg:coupling-mlmc}). 

\subsection{Transport map configurations and proposal mechanisms}\label{subsec: transport-map-configurations-and-proposal-mechanisms}
\begin{figure}[h!]
    \centering
    \begin{subfigure}[b]{0.6\textwidth}
        \centering
        \def\svgwidth{\textwidth}
        %% Creator: Inkscape 1.4.2 (ebf0e940, 2025-05-08), www.inkscape.org
%% Accompanies image file 'synce.pdf' (pdf, eps, ps)
\begingroup%
  \makeatletter%
  \providecommand\color[2][]{%
    \errmessage{(Inkscape) Color is used for the text in Inkscape, but the package 'color.sty' is not loaded}%
    \renewcommand\color[2][]{}%
  }%
  \providecommand\transparent[1]{%
    \errmessage{(Inkscape) Transparency is used (non-zero) for the text in Inkscape, but the package 'transparent.sty' is not loaded}%
    \renewcommand\transparent[1]{}%
  }%
  \providecommand\rotatebox[2]{#2}%
  \newcommand*\fsize{\dimexpr\f@size pt\relax}%
  \newcommand*\lineheight[1]{\fontsize{\fsize}{#1\fsize}\selectfont}%
  \ifx\svgwidth\undefined%
    \setlength{\unitlength}{712.24102783bp}%
    \ifx\svgscale\undefined%
      \relax%
    \else%
      \setlength{\unitlength}{\unitlength * \real{\svgscale}}%
    \fi%
  \else%
    \setlength{\unitlength}{\svgwidth}%
  \fi%
  \global\let\svgwidth\undefined%
  \global\let\svgscale\undefined%
  \makeatother%
  \begin{picture}(1,0.56160764)%
    \lineheight{1}%
    \setlength\tabcolsep{0pt}%
    \put(0,0){\includegraphics[width=\unitlength,page=1]{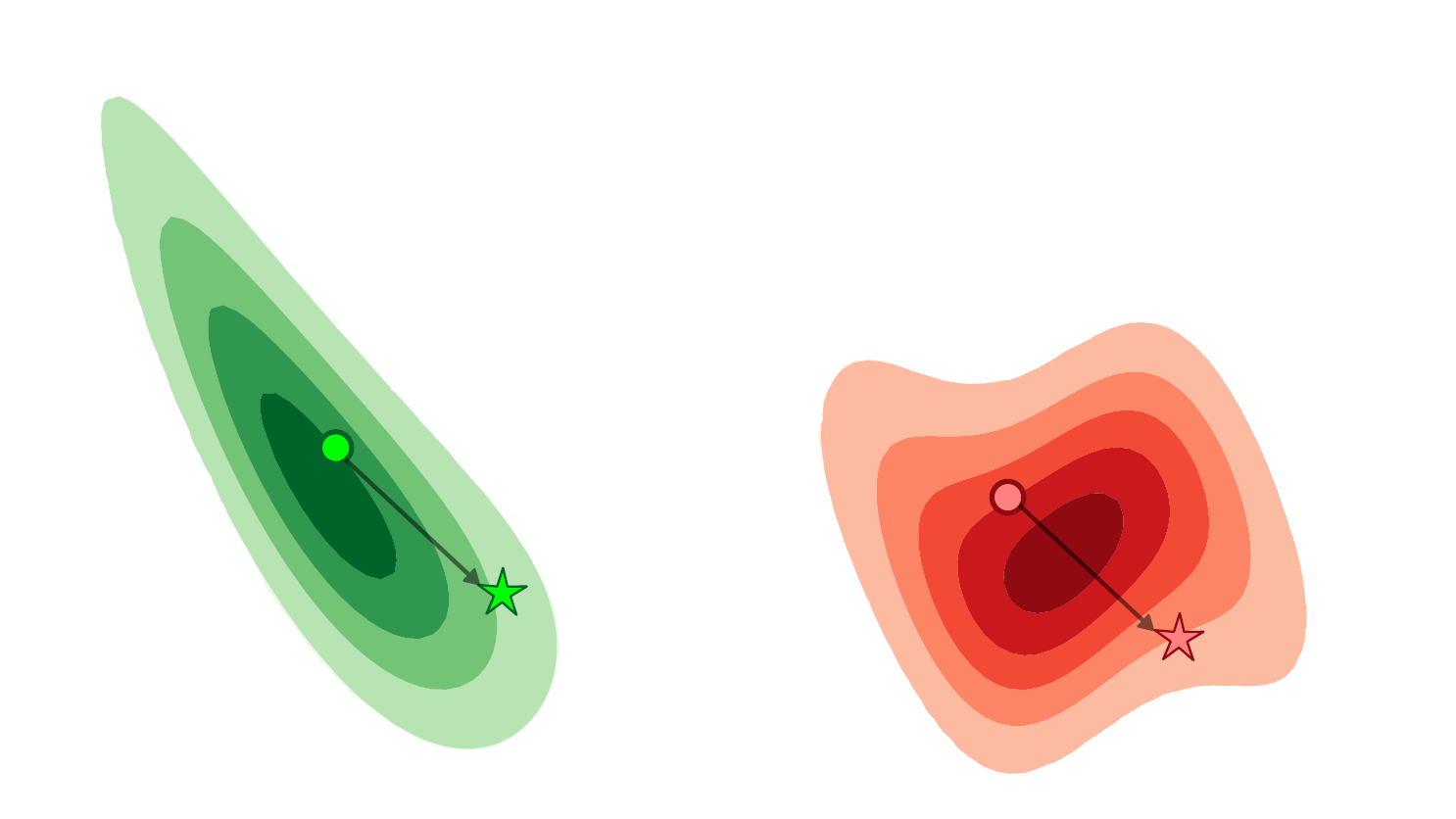}}%
    \put(0.23649902,0.35620924){\color[rgb]{0,0,0}\makebox(0,0)[lt]{\lineheight{1.25}\smash{\begin{tabular}[t]{l}$\pi_{\ell-1}$\end{tabular}}}}%
    \put(0.6615118,0.36084574){\color[rgb]{0,0,0}\makebox(0,0)[lt]{\lineheight{1.25}\smash{\begin{tabular}[t]{l}$\pi_{\ell}$\end{tabular}}}}%
    \put(0.20582979,0.29336434){\color[rgb]{0,0,0}\makebox(0,0)[lt]{\lineheight{1.25}\smash{\begin{tabular}[t]{l}$\theta_{\ell-1}^i$\end{tabular}}}}%
    \put(0.35661427,0.09623003){\color[rgb]{0,0,0}\makebox(0,0)[lt]{\lineheight{1.25}\smash{\begin{tabular}[t]{l}$\theta_{\ell-1}^*$\end{tabular}}}}%
    \put(0.8239668,0.06836552){\color[rgb]{0,0,0}\makebox(0,0)[lt]{\lineheight{1.25}\smash{\begin{tabular}[t]{l}$\theta_{\ell}^*$\end{tabular}}}}%
    \put(0.67518372,0.26353613){\color[rgb]{0,0,0}\makebox(0,0)[lt]{\lineheight{1.25}\smash{\begin{tabular}[t]{l}$\theta_{\ell}^i$\end{tabular}}}}%
  \end{picture}%
\endgroup%

        \caption{SYNCE coupling in the target space.}
        \label{fig: synce}
    \end{subfigure}
  
    \centering
    \begin{subfigure}[b]{0.48\textwidth}
        \centering
        \def\svgwidth{\textwidth}
        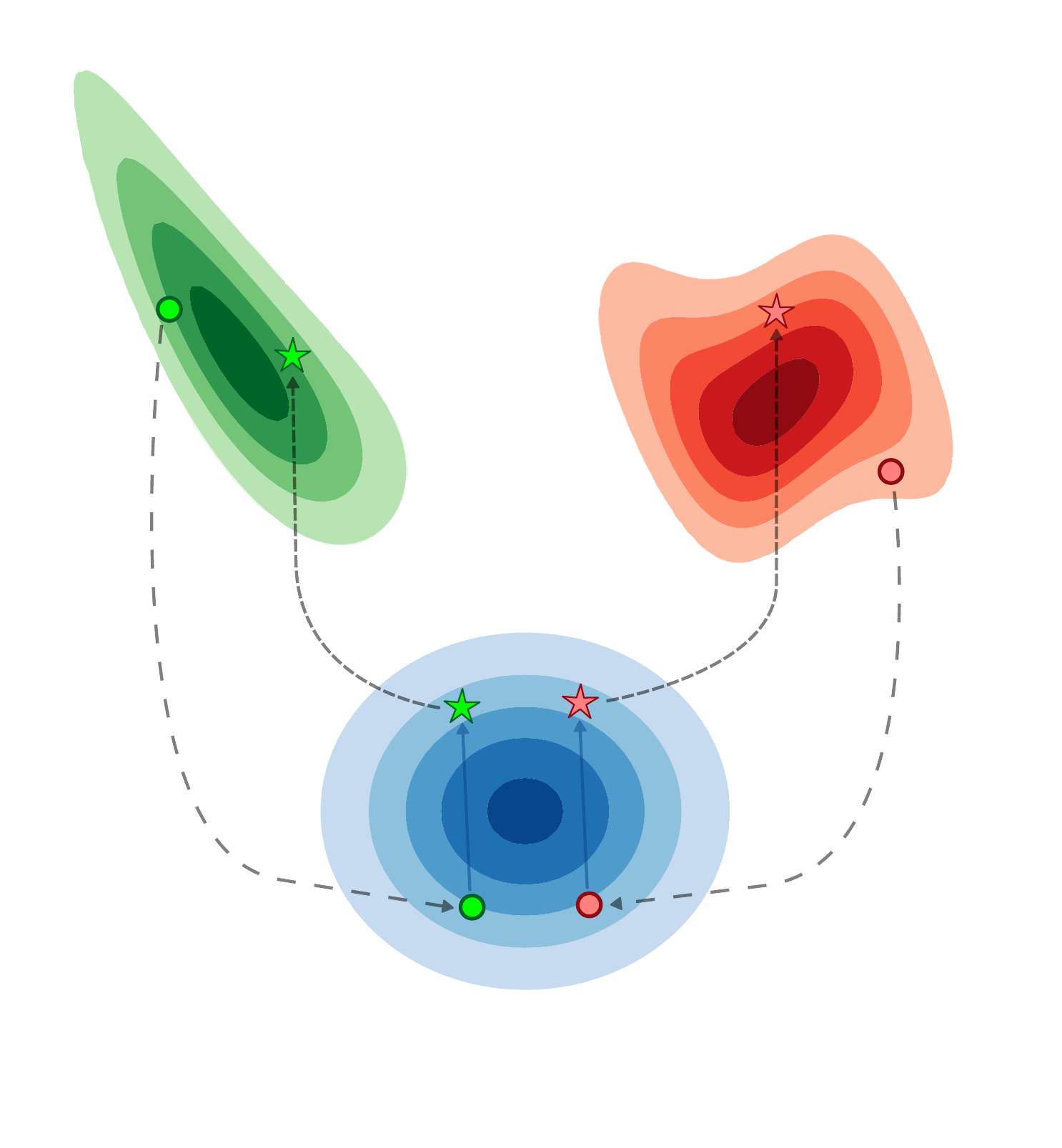
        \caption{TM-SYNCE - random walk direct architecture.}
        \label{fig: rw-direct}
    \end{subfigure}
    \hfill
    \begin{subfigure}[b]{0.48\textwidth}
        \centering
        \def\svgwidth{\textwidth}
        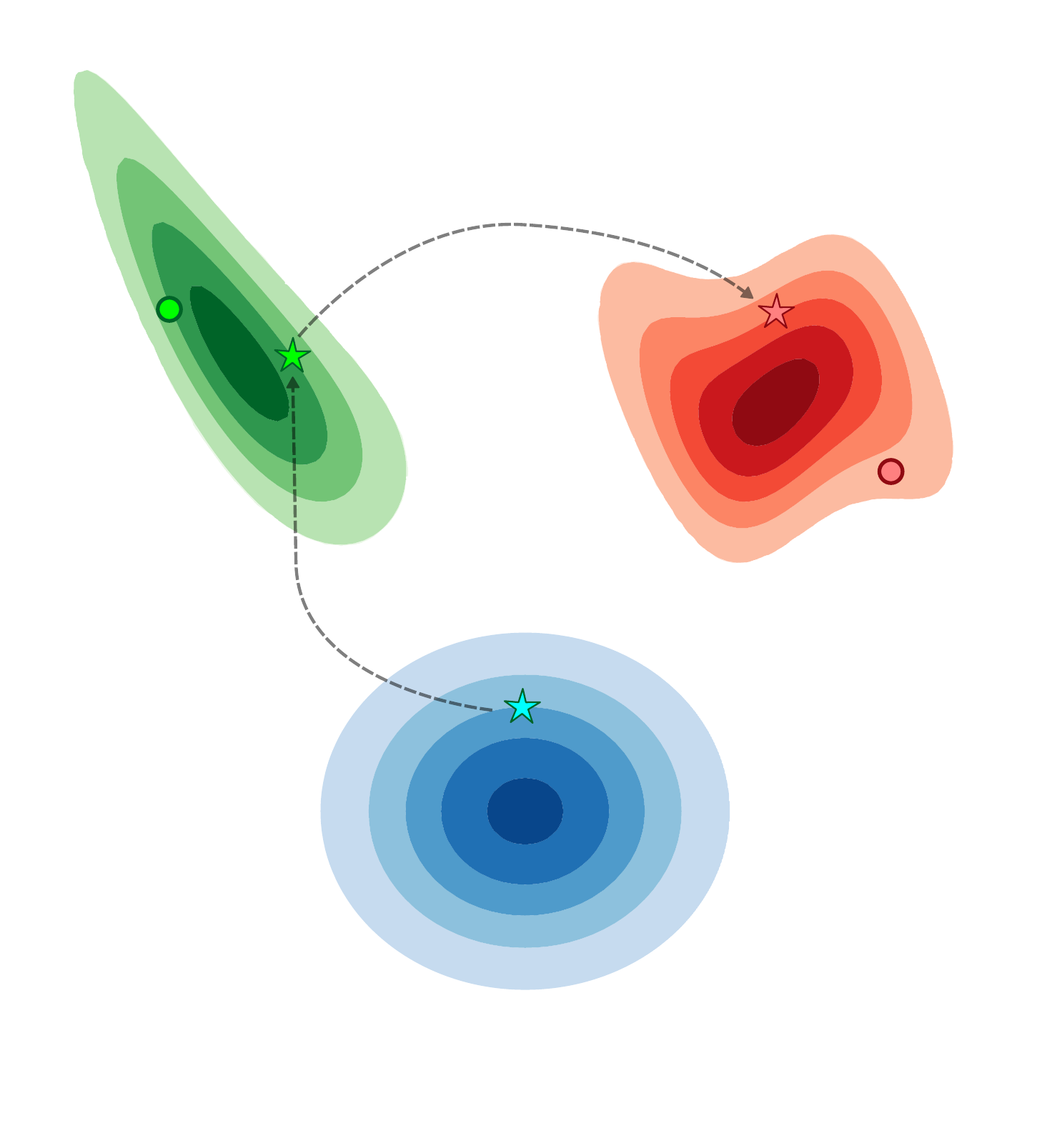
        \caption{TM-SYNCE - independent deep architecture.}
        \label{fig: ind-deep}
    \end{subfigure}
    \caption{The top~\Cref{fig: synce} shows the SYNCE coupling strategy introduced in~\cite{muchandimathSynchronizedStepMultilevel2025} applied directly in the target space, while the bottom~\Cref{fig: rw-direct,fig: ind-deep} illustrate the TM-SYNCE framework we propose in this work that combines transport maps with SYNCE coupling in a shared reference space.~\Cref{fig: rw-direct} depicts the random walk proposal mechanism with direct transport map configuration, and~\Cref{fig: ind-deep} shows the independent proposal mechanism with deep transport map configuration.}
    \label{fig:synce-transport}
\end{figure}

Our framework supports two different ways of learning and applying the transport maps --- direct and deep. In the direct approach, each fidelity level has its own transport map that pushes forward to the same reference distribution ($\pi_{\ell} \leftrightarrow \pi_r$ and $\pi_{\ell-1} \leftrightarrow \pi_r$). This decouples training across fidelities and simplifies deployment since each map is used only with its own level. In the deep approach, the fine level map first pushes the fine posterior to the coarse posterior, and the coarse map then sends both to the reference space ($\pi_{\ell} \leftrightarrow \pi_{\ell-1} \leftrightarrow \pi_r$). The deep approach is particularly useful when the posterior distributions of adjacent fidelity levels are similar because it leverages the structure of the coarse posterior to inform the mapping of the fine. This can lead to more efficient sampling and better correlation between the proposals. We show the two proposal mechanisms with two different map configurations along with the original SYNCE coupling in~\Cref{fig:synce-transport}. All four combinations of map configurations (direct, deep) and proposal mechanisms (random walk, independent) are supported in our framework and we show two of the combinations in the figure for clarity.

\subsection{Implementation details}\label{subsec: implementation-details}
To implement the TM-SYNCE framework, we require transport maps $T_\ell$ and $T_{\ell-1}$ that approximate the pushforward conditions $T_{\ell \#} \mu_\ell \approx \mu_r$ and $T_{(\ell-1) \#} \mu_{\ell-1} \approx \mu_r$ for the direct configuration (or $T_{\ell \#} \mu_\ell \approx \mu_{\ell-1}$ and $T_{(\ell-1) \#} \mu_{\ell-1} \approx \mu_r$ for the deep configuration), where $\mu_{\ell}, \mu_{\ell-1}$ and $\mu_r$ are the measures associated with densities $\pi_{\ell}, \pi_{\ell-1}$ and $\pi_r$ respectively. In this work, we choose the reference distribution to be a standard multivariate Gaussian $\pi_r = \mathcal{N}(\bm{0}, \bm{I})$. We choose the transport map architectures to ensure fast evaluation of forward/inverse mappings and Jacobian determinants, and we train the maps using MCMC samples from the target distributions. The following subsections provide details on the map architecture selection and training process.

\subsubsection{Map architecture selection}\label{subsubsec: map-architecture-selection}
We explore three popular transport map architectures that balance expressiveness and computational efficiency required for MCMC sampling:
\begin{enumerate}
    \item Optimal transport and Brenier maps: Optimal transports are a natural choice for transport maps and are obtained by minimizing a cost function between distributions through the Kantorovich problem~\cite{villaniOptimalTransportOld}. When the cost is quadratic, Brenier's theorem~\cite{brenierPolarFactorizationMonotone1991} guarantees that the unique optimal transport map takes the form $T = \nabla \psi$ where $\psi$ is a convex potential. This map is particularly attractive because it aligns the map with the directions of \textit{maximal correlation} between the source and target distributions, making it the most efficient coupling strategy. For Gaussian measures $\mu = \mathcal{N}\left(m_{\mu},\Sigma_{\mu}\right)$ and $\nu = \mathcal{N}\left(m_{\nu},\Sigma_{\nu}\right)$, the Brenier map is given by the closed-form expression $T_{\text{Brenier}}(x) = m_{\nu} + A\left(x - m_{\mu}\right)$, where $A = \Sigma_{\mu}^{-1/2}\left(\Sigma_{\mu}^{1/2}\Sigma_{\nu}\Sigma_{\mu}^{1/2}\right)^{1/2}\Sigma_{\mu}^{-1/2}$~\cite{knottOptimalMappingDistributions1984} is a linear transformation that aligns the covariance matrices of the two distributions. This optimal correlation property is precisely what we need for multi-fidelity coupling, however, analytical closed form solutions for Brenier maps do not exist except for Gaussian, elliptical and some simple measures~\cite{knottOptimalMappingDistributions1984,olkinDistanceTwoRandom1982,villaniOptimalTransportOld}.

    \item Lower triangular maps: Lower triangular (LT) maps offer an alternative to Brenier maps through weighted quadratic costs. When the cost function $c(x,y)$ in the Kantorovich problem is chosen to be a weighted quadratic of the form $c(x,y) = \sum_{i=1}^{d} t_i\left|x_i-y_i\right|^2, t_i > 0$, the optimal transport map converges to the Knothe-Rosenblatt (KR) map as the weights dominate one another (specifically, as $t_i \rightarrow 0$)~\cite{bonnotteKnothesRearrangementBreniers2013}. In practice, lower triangular maps are used as approximations to the computationally intractable Brenier maps. Each component of a lower triangular map depends only on the current and preceding variables, which makes both the forward and inverse mappings, as well as the computation of the Jacobian determinant, highly efficient. For these reasons, LT maps are favoured for MCMC sampling~\cite{parnoTransportMapAccelerated2018,peherstorferTransportbasedMultifidelityPreconditioner2019}. 

    \item Normalizing flows: Posteriors arising from inverse problems are often complex, non-Gaussian and sometimes multi-modal, making it challenging to capture their structure using simple transport maps. Normalizing flows (NF)~\cite{papamakariosNormalizingFlowsProbabilistic,dinhDensityEstimationUsing2017} address this challenge by constructing neural network-parameterized sequences of invertible transformations that can represent highly complex, nonlinear mappings between reference and target distributions. Their compositional structure maintains computational efficiency for forward/inverse evaluation and Jacobian determinants, while leveraging automatic differentiation and GPU acceleration for efficient training.
\end{enumerate}

\subsubsection{Map training}\label{subsubsec: map-training}
We train the parameters $\bm{\phi}$ of the chosen transport map architecture $T\left(\cdot; \bm{\phi}\right)$ using MCMC samples of the target distribution $\left\{\bm{\theta}^1, \bm{\theta}^1, \ldots, \bm{\theta}^{N} \right\} \sim \pi$. The training process optimizes the map to minimize the discrepancy between the target distribution and the pullback of the target distribution through the map~\cite{parnoTransportMapAccelerated2018,papamakariosNormalizingFlowsProbabilistic}. The choice of loss function is crucial and depends on both the map architecture and the application context. The most common loss metric used is the forward Kullback--Leibler (KL) divergence~\cite{kullbackInformationSufficiency1951}, defined as
\begin{align}\label{eq: forward-kl}
    D_{KL}\left(\pi \| \tilde{\pi}\right) &= \mathbb{E}_{\pi}\left[\log\left(\frac{\pi(\bm{\theta})}{\tilde{\pi}(\bm{\theta})}\right)\right], \\ \nonumber
    &= \mathbb{E}_{\pi}\left[\log \pi(\bm{\theta}) - \log \pi_r\left(T(\bm{\theta}; \bm{\phi})\right) - \log \left|\det J_{T}(\bm{\theta})\right|\right],
\end{align}
where $J_T$ is the Jacobian of the transport map $T$. The equation above is obtained by using the change of variables formula $\tilde{\pi}(\bm{\theta}) = \pi_r(T\left(\bm{\theta};\bm{\phi}\right))\left|\det J_T(\bm{\theta})\right|$. The loss function can then be written as
\begin{align}\label{eq: KL-loss}
    \mathbf{\mathcal{L}}_{KL}(\bm{\theta}; \bm{\phi}) = -\mathbb{E}_{\pi}\left[\log \pi_r\left(T(\bm{\theta}; \bm{\phi})\right) + \log \left|\det J_{T}(\bm{\theta}; \bm{\phi})\right| \right].
\end{align}
We estimate the expectation in the above equation empirically using MCMC samples from the target distribution via sample average approximation~\cite{kleywegt2002sample} as
\begin{align}\label{eq: SAA-loss}
    \mathbf{\hat{\mathcal{L}}}(\bm{\phi}; \mathcal{S}_n) = \frac{1}{|\mathcal{S}_n|}\sum_{\bm{\theta}^i \in \mathcal{S}_n}\mathbf{\mathcal{L}}_{KL}(\bm{\theta}^i; \bm{\phi}),
\end{align}
where $\mathcal{S}_n \subseteq \{\bm{\theta}^1, \bm{\theta}^2, \ldots, \bm{\theta}^{n}\}$ denotes a subset of the first $n$ MCMC samples, $|\mathcal{S}_n|$ denotes the cardinality of set $\mathcal{S}_n$, and, $\mathbf{\mathcal{L}}_{KL}\left(\bm{\theta}^i; \bm{\phi}\right)$ is the loss evaluated at the $i$'th sample using~\Cref{eq: KL-loss}. In practice, we optimize the transport map parameters $\bm{\phi}$ periodically every $K$ MCMC iterations to adapt the map to the evolving sample distribution~\cite{parnoTransportMapAccelerated2018}. The choice of the optimization objective, sample set $S_{jK}$ where $jK=i$ $\forall j\in \mathbb{N}$, and optimization algorithm depend on the map architecture. We summarize the periodic optimization step for both LT maps and NF below in~\Cref{alg:optimization}.

\begin{algorithm}[ht]
    \caption{Periodic transport map optimization}\label{alg:optimization}
    \begin{algorithmic}[1]
        \STATE \textbf{Input:} 
        \STATE \hspace{1em} Previous map parameters $\bm{\phi}^{(i-1)}$
        \STATE \hspace{1em} Current MCMC samples $\{\bm{\theta}^1, \bm{\theta}^2, \ldots, \bm{\theta}^{i}\}$
        \STATE \hspace{1em} Optimization frequency $K$
        \vspace{1em}
        \STATE \textbf{Output:} 
        \STATE \hspace{1em} Updated map parameters $\bm{\phi}^{(i)}$
        \vspace{1em}
        \STATE \textbf{Step 1: Determine if optimization is needed}
        \STATE \textbf{if} $(i \mod K) = 0$ \textbf{then} \hfill \textit{(optimize every K iterations)}
        \vspace{1em}
        \STATE \hspace{1em} \textbf{Step 2: Select sample set $\mathcal{S}_{i}$}
        \STATE \hspace{1em} Set $\mathcal{S}_i = \begin{cases}
            \{\bm{\theta}^1, \bm{\theta}^2, \ldots, \bm{\theta}^{i}\} & \text{ (full batch for LT maps)} \\
            \{\bm{\theta}^{i-M}, \bm{\theta}^{i-M+1}, \ldots, \bm{\theta}^{i}\} & \text{ (sliding window M for NF)}
        \end{cases}$
        \vspace{1em}
        \STATE \hspace{1em} \textbf{Step 3: Optimize map parameters}
        \STATE \hspace{1em} Compute $\bm{\phi}^{(i)} = \begin{cases}
            \arg \min_{\bm{\phi}} \mathbf{\hat{\mathcal{L}}}(\bm{\phi}; \mathcal{S}_{i}) & \text{ (deterministic optimization for LT maps)} \\
            \bm{\phi}^{(i-1)} - \alpha \nabla_{\phi}\mathbf{\hat{\mathcal{L}}}(\bm{\phi}; \mathcal{S}_i) & \text{ (noisy optimization for NF)}
        \end{cases}$
        \STATE \hspace{1em} where $\alpha$ is the learning rate
        \STATE \textbf{else}
        \STATE \hspace{1em} Set $\bm{\phi}^{(i)} = \bm{\phi}^{(i-1)}$ \hfill \textit{(no optimization)}
        \STATE \textbf{end if}
        \vspace{1em}
        \RETURN $\bm{\phi}^{(i)}$
    \end{algorithmic}
\end{algorithm}

LT maps use full-batch, deterministic optimization because the objective function in~\Cref{eq: KL-loss} is convex and separable for this architecture~\cite{parnoTransportMapAccelerated2018}. We warm-start the optimization using the previous map parameters $\bm{\phi}^{(i-1)}$ and use a Newton method to solve the optimization problem. The comprehensive dataset helps capture the full complexity of the target distribution but can be computationally expensive as the number of samples grows. 

NF use mini-batch, noisy optimization because the objective function in~\Cref{eq: KL-loss} is non-convex for neural parameterizations~\cite{papamakariosNormalizingFlowsProbabilistic}. The optimization is performed using stochastic gradient descent (SGD) with the Adam optimizer~\cite{kingma2015adam}. To manage computational cost, we use a sliding window of the most recent $M$ samples for training. 

Multi-level settings offer additional opportunities for efficiency. Because the coarse posterior at MLMC level $\ell$ coincides with the fine posterior at level $\ell-1$, we initialize the current coarse map $T_{\ell-1}$ using the previously trained fine map $T_{\ell}$ from the prior MLMC level. This bootstrap approach is particularly beneficial when the fidelity levels are closely related, allowing us to leverage existing knowledge about the posterior structure.

\subsubsection{Reference covariance selection}\label{subsubsec: reference-covariance-selection}
Beyond training the transport maps, we also need to select the reference covariance $\bm{C}_r$ used in the SYNCE coupling strategy in~\Cref{eq: synce-rw,eq: synce-ind}. The choice of $\bm{C}_r$ significantly impacts the performance of the coupled MCMC algorithm, as it controls both the step size and correlation structure of the proposals. A step size that is too small can lead to slow exploration of the posterior, while a step size that is too large can result in low acceptance rates. Since the transport maps Gaussianize the target distributions, we use the optimal scaling rule for Gaussian targets~\cite{andrieuTutorialAdaptiveMCMC2008} $\bm{C}_r = 2.38^2 / d \times \bm{I}_d$, where $d$ is the dimension of the parameter space. We also adapt the scaling of $\bm{C}_r$ using popular adaptive MCMC strategies~\cite{haarioAdaptiveMetropolisAlgorithm2001,andrieuTutorialAdaptiveMCMC2008} by tailoring the proposal distribution to the local geometry of the reference distribution.

\section{Results}\label{sec: results}
In this section, we present the results of our proposed framework on a simple test problem and on a more complex turbulence parameter estimation problem. The simple test problem involves two two-dimensional posterior distributions --- a banana distribution and a quartic distribution. We use this problem to demonstrate the effectiveness of our proposed framework involving coupled reference proposals and transport maps in achieving correlation between the samples while maintaining good mixing properties. We also compare the performance of different transport map architectures and configurations to the true map results. The second problem involves a more complex application of our framework to a turbulence parameter estimation problem for computational fluid dynamics (CFD) simulations. We show how our framework can be used to effectively reduce the variance of the output estimates by using a surrogate with our proposed multi-fidelity coupling strategy.

We explore two different types of transport maps in our experiments --- lower triangular (LT) maps and normalizing flows (NF). The lower triangular maps are implemented using the monotone parameterization toolkit \texttt{MParT} library~\cite{mpart2022}, a high-performance library designed for parameterizing and evaluating monotone lower triangular transport maps. The normalizing flows are implemented using the \texttt{normflows} package~\cite{Stimper2023}, a PyTorch-based library for constructing and training normalizing flows. This package offers the flexibility to experiment with various flow architectures and is well-suited for deep learning applications. We do not consider Brenier maps, as general-purpose implementations for arbitrary continuous distributions are currently unavailable, and this remains an active area of research~\cite{feydy2019interpolating,muzellec2019subspace}. Moreover, evaluating the Jacobian in such formulations is computationally challenging because of dense maps, further limiting their practical use in our setting.

\subsection{Banana and quartic distributions}\label{subsec: banana-quartic}
\begin{figure}[ht]
    \centering
    \includegraphics[width=0.75\textwidth]{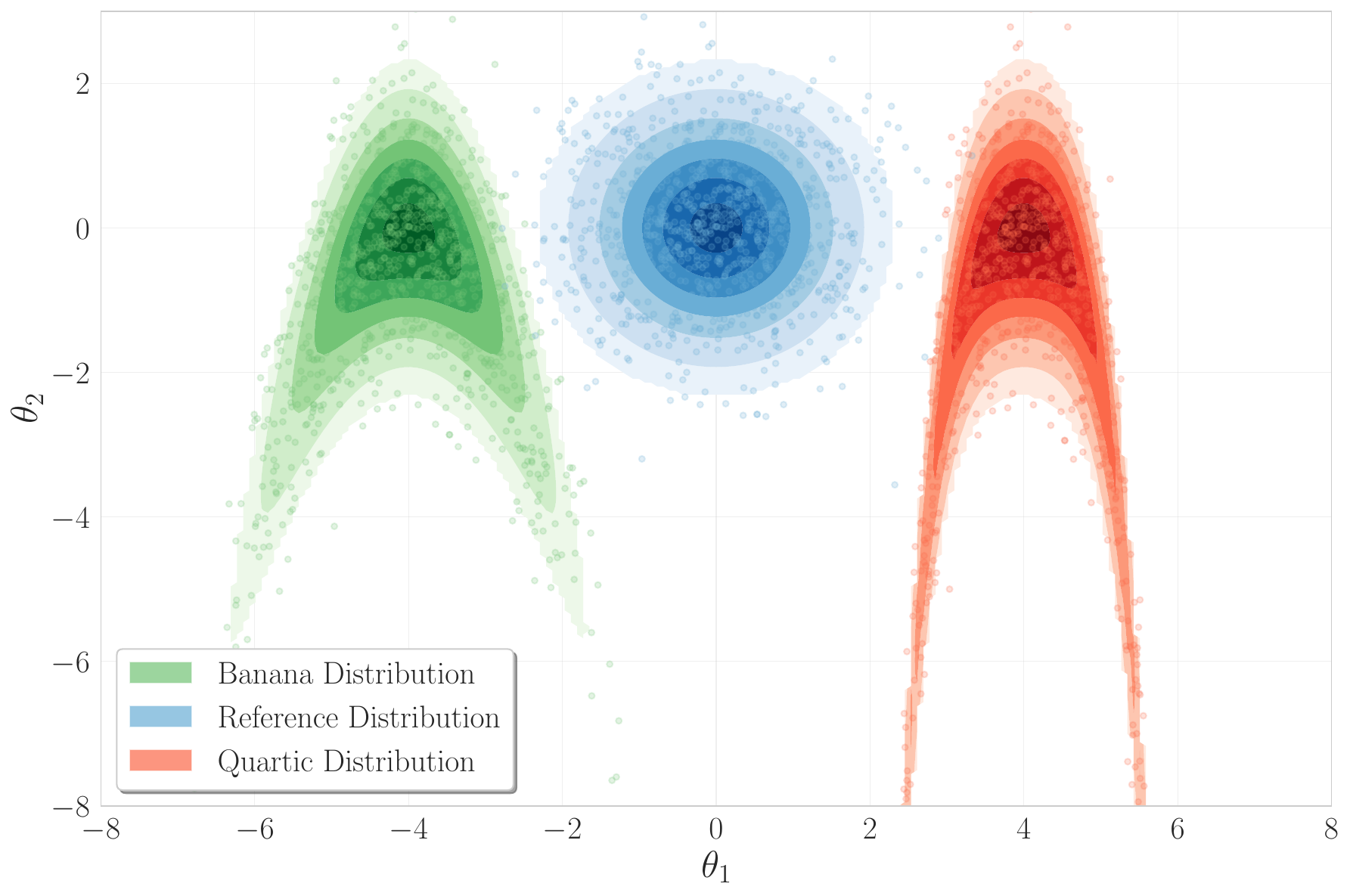}
    \caption{Banana and quartic distributions. The standard Gaussian distribution is also plotted for reference. Samples from the banana and quartic distributions are shown in green and red, respectively. These samples are generated using the analytical transport maps in~\Cref{eq: banana-quartic-transport} from the reference Gaussian samples. Note that the posteriors are dissimilar with high curvature, making it challenging to generate correlated samples while maintaining good mixing properties.}
    \label{fig: banana-quartic-models}
\end{figure}

The banana and the quartic distributions are two well-known test problems in the MCMC literature. Samples from the two distributions can be generated by using analytical lower triangular transformations from a reference Gaussian distribution. The banana and quartic distributions are defined as:
\begin{align}\label{eq: banana-quartoc-distribution}
    \pi_{\text{banana}}(\bm{x}) &\propto \mathcal{N}\left(x_1 - s_1\right)\times \mathcal{N}\left(x_2 + (x_1 -s_1)^2\right), \\ \nonumber
    \pi_{\text{quartic}}(\bm{x}) &\propto \mathcal{N}\left(x_1 - s_2\right)\times \mathcal{N}\left(x_2 + (x_1 -s_2)^2 + (x_1-s_2)^4\right),
\end{align}
where $s_1 = -4.0$ and $s_2 = 4.0$ are the shift parameters for the two distributions. The shift parameters are included to ensure that the two distributions do not overlap significantly, making the problem more challenging for generating correlated samples. The contours and samples from the two distributions are shown in~\Cref{fig: banana-quartic-models}. The exact transport maps to the reference Gaussian space are given as (note the direction of the transport maps from the target space to the reference space)
\begin{align}\label{eq: banana-quartic-transport}
    T_{\text{banana}}(x_1, x_2) &= \begin{bmatrix}
        x_1 - s_1, \\
        x_2 + (x_1 - s_1)^2
    \end{bmatrix}, \quad T_{\text{quartic}}(x_1, x_2) = \begin{bmatrix}
        x_1 - s_2, \\
        x_2 + (x_1 - s_2)^2 + (x_1 - s_2)^4
    \end{bmatrix}.
\end{align}
In the context of deep adaptation, we define an analytical transport map from the quartic distribution to the banana distribution as
\begin{align}\label{eq: banana-quartic-deep-transport}
    T_{\text{quartic} \rightarrow \text{banana}}(x_1, x_2) &= \begin{bmatrix}
        x_1 + s_1 - s_2, \\
        x_2 + (x_1 - s_2)^4
    \end{bmatrix}.
\end{align} 

In this section, we assume that the banana distribution is the coarse (low-fidelity) posterior and the quartic distribution is the fine (high-fidelity) posterior. We compare the performance of different multi-fidelity coupling strategies --- (1) the SYNCE framework without transport maps applied directly on the target space~\cite{muchandimathSynchronizedStepMultilevel2025}, (2) the TM-SYNCE framework with lower triangular transport maps~\cite{parnoTransportMapAccelerated2018,mpart2022}, (3) the TM-SYNCE framework with normalizing flows~\cite{papamakariosNormalizingFlowsProbabilistic}, and (4) the TM-SYNCE framework with the true maps in~\Cref{eq: banana-quartic-transport}. We also compare the performance of different configurations -- direct and deep with two different resynchronization parameters $\omega = 0.0$ and $0.5$. 

For the lower triangular transport maps, \texttt{MParT}~\cite{mpart2022} employs polynomial-based monotonic triangular functions with adaptive polynomial orders based on the complexity of each distribution. We use total order 2 polynomials for the banana distribution to capture the quadratic non-linearity, and, use total order 4 polynomials for the quartic distribution. In the deep architecture, we use a total order 3 for the fine to coarse level mapping combined with the coarse model's gradient information, enabling efficient composition of transformations. For the normalizing flow implementation using \texttt{normflows}~\cite{Stimper2023}, we employ Real NVP~\cite{dinhDensityEstimationUsing2017} transformations. The flows consist of multiple affine coupling blocks using multi-layer perceptrons (MLPs). Specifically, we use 8 coupling layers with 3 hidden layers of 16 units each for the coarse distribution, while the fine distribution employs 16 coupling layers with 3 hidden layers of 32 units each to capture the more complex quartic geometry. Each coupling block is followed by coordinate permutation layers that swaps coordinates to ensure all dimensions are transformed.

For the SYNCE framework without transport maps, we sample $\bm{\eta} \sim \mathcal{N}\left(\bm{0}, \bm{I}_2\right)$ and scale $\bm{\eta}$ by $2.38^2 / 2 \times \sqrt{\bm{\Sigma}_{\text{Laplace}}}$, where $\bm{\Sigma}_{\text{Laplace}}$ is the covariance matrix obtained by Laplace approximation of the respective posterior distribution. For the SYNCE framework with transport maps, we use a reference space proposal covariance $\bm{C}_r = 2.38^2/2 \times \bm{I}_2$. The transport maps are trained every $K = 5000$ MCMC iterations, with $K$ defined in~\Cref{alg:optimization}. For lower triangular transport maps, we optimize the map parameters using all the samples collected so far, with \texttt{scipy.minimize}'s default BFGS optimizer. For normalizing flows, we use only the recent $M = 5000$ samples to compute the loss and optimize the map parameters with the Adam optimizer~\cite{kingma2015adam}. We use batches of 1024 samples over 200 epochs and learning rates of $1 \times 10^{-3}$ and $1 \times 10^{-4}$ for the banana and quartic distributions, respectively.

We choose the Pearson correlation coefficient ($\rho$), the integrated autocorrelation time ($\tau$), the effective sample size (ESS), and a normalized KL divergence variance of the form $\text{KL}_m = 1 - \exp(-\sigma^2_M)$, where $\sigma^2_M = \mathbb{V}\text{ar}\left[\log \pi(\bm{\theta}) - \log \pi_r(T(\bm{\theta})) - \log \left|\det J_T (\bm{\theta})\right| \right]$~\cite{parnoTransportMapAccelerated2018} as metrics to evaluate the performance of the different transport map architectures and adaptation strategies. The Pearson correlation coefficient describes the correlation between the samples from the two posteriors, with higher values indicating better coupling and greater variance reduction. The ESS and the integrated autocorrelation time (IAT) are used to measure the mixing properties of the samples, with higher ESS and lower IAT indicating better mixing. The KL divergence variance is used to measure the quality of the transport maps, with lower values indicating better maps that capture the geometry of the target posterior distributions.

\begin{figure}[ht]
    \centering
    \includegraphics[width=\textwidth]{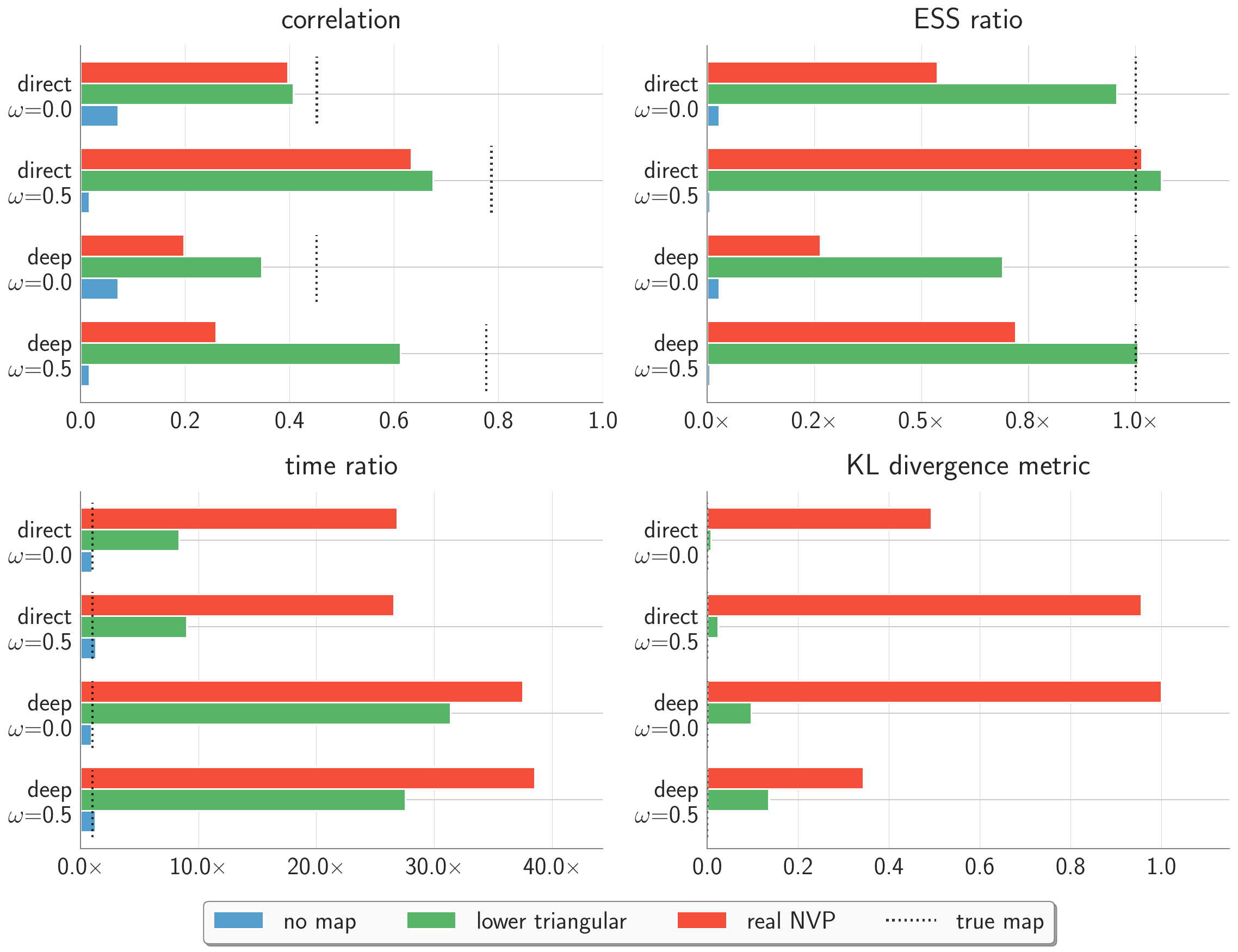}
    \caption{Relative comparison of different multi-fidelity coupling methods on the banana and quartic distributions. The plots show the worst case (over all trials) Pearson correlation coefficient ($\rho$), effective sample size (ESS), total time taken and KL divergence metric ($\sigma^2_M$) across the two dimensions for each method. TM-SYNCE with resynchronization and lower triangular maps in the direct configuration performs the best across all metrics.}
    \label{fig: banana-quartic-comparison}
\end{figure}

For all architectures, we run 5 independent chains of 100,000 MCMC iterations each, with the first 30,000 samples discarded as burn-in. For each metric apart from the total time, we take the worse value over the two dimensions and report the median value across the 5 independent chains. For the total time, we report the mean value across the 5 independent chains. We summarize the results in~\Cref{fig: banana-quartic-comparison} and~\Cref{tab:sampling_performance}.

Figure~\ref{fig: banana-quartic-comparison} demonstrates the effectiveness of transport map architectures across multiple performance metrics. The top-left plot shows the Pearson correlation coefficient in~\Cref{eq: mlmc-variance} of different architectures and the true map. The results indicate that both lower triangular and real NVP transport maps significantly enhance the correlation between samples compared to the baseline SYNCE framework without transport maps. Correlation is further improved by resynchronizing the chains (setting $\omega = 0.5$), a result that is consistent with our previous work~\cite{muchandimathSynchronizedStepMultilevel2025}. Notably, the direct architectures for both lower triangular and real NVP maps achieve the highest correlation, with the direct lower triangular map performing slightly better than the direct real NVP map. 

The top-right plot shows the effective sample size (ESS) ratio compared to the true map. The results indicate that both lower triangular and real NVP transport maps significantly improve the ESS compared to the baseline SYNCE framework without transport maps. Surprisingly, the direct architectures for both lower triangular and real NVP maps with resynchronization surpass the true map performance, a result that we attribute to the approximate maps smoothing out some of the complexities of the target distributions. This smoothing effect facilitates improved chain mixing and reduced autocorrelation between samples. Similar to the correlation results, the direct architectures for both lower triangular and real NVP maps achieve the highest ESS, with the direct lower triangular map performing slightly better than the direct real NVP map. 

The bottom-left plot shows the time ratio compared to the true map. There are two key observations from this result. First, the direct architectures for both the maps are more efficient than the deep architectures. This is attributed to the fact that optimization for lower triangular maps in deep architectures loses the convex and separable properties that make direct architectures computationally tractable~\cite{parnoTransportMapAccelerated2018}. Second, the deep real NVP architecture is significantly more expensive than all other architectures. This is likely due to the architectural complexity of neural networks in capturing curvy distributions and the challenges posed by limited training data.

The bottom right plot shows the KL divergence metric with the true map value equal to 0. The no map approach does not have a corresponding KL divergence value since there is no transport map involved. The results indicate that neural networks struggle to capture the banana and quartic posteriors, leading to high KL divergence values. We attribute this to the true map being lower triangular, which the real NVP architecture struggles to capture under limited training data. Overall, the results favour transport map integration to achieve orders-of-magnitude improvements in sampling quality and efficiency, but also suggest that architectural choice should be closely aligned with practical demands for computational speed.

While the relative comparisons in~\Cref{fig: banana-quartic-comparison} provide valuable insights into method performance, the absolute metrics in~\Cref{tab:sampling_performance} reveal the substantial improvements achieved by the proposed framework. Both LT and real NVP maps, when used with the direct architecture, show dramatic gains—achieving up to $39\times$ and $37\times$ improvement in worst-case Pearson correlation and $24\times$ and $7.7\times$ increase in effective sample size per second, respectively, over the baseline SYNCE (no map) method at $\omega=0.5$. The true map delivers the best performance because it does not incur the computational cost of training or evaluation. However, the absolute ESS for the learned LT and NVP maps closely matches that of the true map, demonstrating that practical, data-driven map constructions are nearly as efficient for sampling in real applications. The choice between direct and deep architectures should be guided by the specific trade-offs between computational efficiency and distributional accuracy requirements, with direct architectures offering superior computational performance, for this particular test case.

\begin{table}[ht]
    \centering
     \caption{Performance comparison of different multi-fidelity coupling methods on the banana and quartic distributions. Transport maps significantly improve the correlation and effective sample size compared to the baseline SYNCE framework without transport maps. Direct LT and NVP with resynchronization perform the best (and are bolded). }
    \label{tab:sampling_performance}
    \begin{tabular}{lccrrrrrr}
        \hline
        Method & $\omega$ & $\rho_{\min}$ & $\tau_{\max}$ & ESS & time & ESS/sec & rel. $\rho_{\min}$ & rel. ESS/sec \\
        \hline
        \multirow{2}{*}{No map} & 0.00 & 0.07 & 277.86 & 252.00 & 11.98 & 21.04 & 1.00 & 1.00 \\
        & 0.50 & 0.02 & 666.04 & 106.00 & 15.94 & 6.65 & 1.00 & 1.00 \\
        \hline
        \multirow{2}{*}{True direct} & 0.00 & 0.45 & 7.87 & 8890.00 & 12.35 & 720.00 & 6.27 & 34.23 \\
        & 0.50 & 0.79 & 4.22 & 16603.00 & 12.41 & 1338.13 & 45.73 & 201.23 \\
        \multirow{2}{*}{True deep} & 0.00 & 0.45 & 7.78 & 9004.00 & 12.70 & 708.96 & 6.26 & 33.70 \\
        & 0.50 & 0.78 & 4.09 & 17133.00 & 12.63 & 1356.21 & 45.16 & 203.95 \\
        \hline
        \multirow{2}{*}{LT direct} & 0.00 & 0.41 & 8.24 & 8501.00 & 103.18 & 82.39 & 5.66 & 3.92 \\
        & 0.50 & 0.68 & 3.98 & 17600.00 & 111.59 & 157.73 & \textbf{39.27} & \textbf{23.72} \\
        \multirow{2}{*}{LT deep} & 0.00 & 0.35 & 11.26 & 6215.00 & 398.46 & 15.60 & 4.81 & 0.74 \\
        & 0.50 & 0.61 & 4.06 & 17232.00 & 347.79 & 49.55 & 35.64 & 7.45 \\
        \hline
        \multirow{2}{*}{NVP direct} & 0.00 & 0.40 & 14.64 & 4781.00 & 331.74 & 14.41 & 5.51 & 0.69 \\
        & 0.50 & 0.63 & 4.16 & 16843.50 & 329.87 & 51.06 & \textbf{36.81} & \textbf{7.68} \\
        \multirow{2}{*}{NVP deep} & 0.00 & 0.20 & 29.47 & 2376.00 & 476.35 & 4.99 & 2.74 & 0.24 \\
        & 0.50 & 0.26 & 5.67 & 12339.50 & 486.52 & 25.36 & 15.07 & 3.81 \\
        \hline
    \end{tabular}
    \begin{flushleft}
    \footnotesize
    The table shows the worst case Pearson correlation coefficient ($\rho_{\min}$), maximum IAT ($\tau_{\max}$), minimum ESS, total time taken (s), ESS per second, relative Pearson correlation coefficient, and relative ESS per second across the two dimensions for each method. The relative values are computed with respect to the SYNCE framework without transport maps.
    \end{flushleft}
\end{table}

\begin{figure}[ht]
    \centering
    \begin{subfigure}[b]{0.48\textwidth}
        \centering
        \def\svgwidth{\textwidth}
        \includegraphics[width=\textwidth]{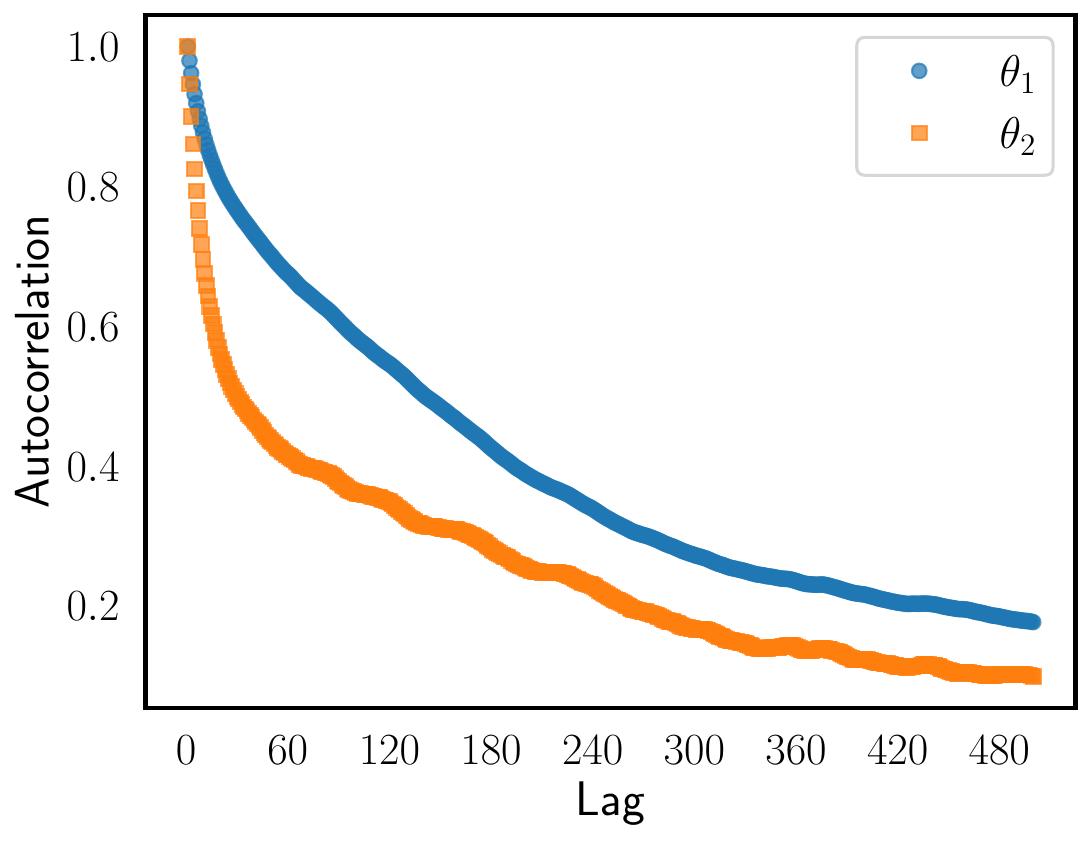}
        \caption{Autocorrelation plot with samples obtained from the multi-fidelity preconditioned MCMC framework~\cite{peherstorferTransportbasedMultifidelityPreconditioner2019}. Effective sample size is 139.}
        \label{fig: precondition-lag}
    \end{subfigure}
    \hfill
    \begin{subfigure}[b]{0.48\textwidth}
        \centering
        \def\svgwidth{\textwidth}
        \includegraphics[width=\textwidth]{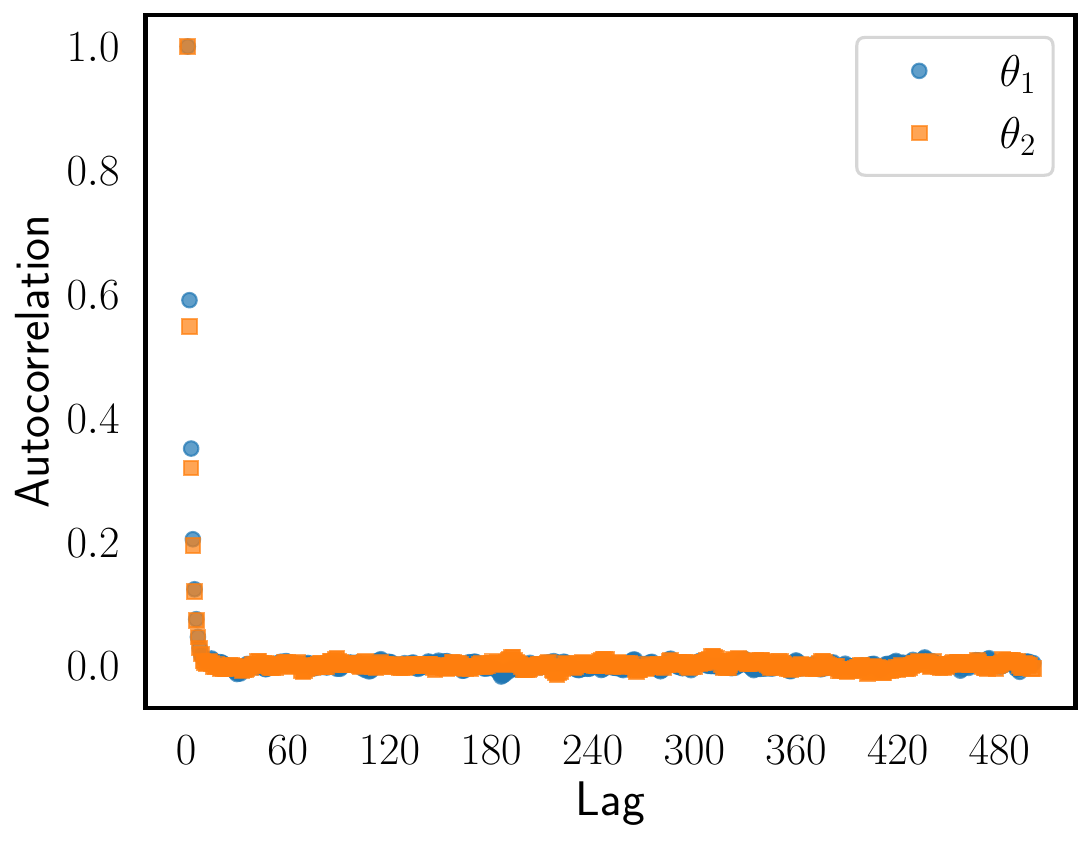}
        \caption{Autocorrelation plot with samples obtained from the TM-SYNCE with lower triangular maps and direct configuration. Effective sample size is 17393.}
        \label{fig: tm-synce-lag}
    \end{subfigure}
    \caption{Autocorrelation comparison between (a) multi-fidelity preconditioned MCMC framework~\cite{peherstorferTransportbasedMultifidelityPreconditioner2019} and (b) TM-SYNCE. The banana distribution is not a good approximation of the quartic distribution, leading to poor mixing in (a). In contrast, TM-SYNCE effectively leverages transport maps to achieve good mixing properties even when the coarse and fine distributions are dissimilar.}
    \label{fig: preconditioned-lag-tm-synce}
\end{figure}

To further illustrate the advantages of our proposed TM-SYNCE framework, we compare the autocorrelation plots of samples obtained from the multi-fidelity preconditioned MCMC framework~\cite{peherstorferTransportbasedMultifidelityPreconditioner2019} and our TM-SYNCE framework with lower triangular maps in the direct configuration. The results are shown in~\Cref{fig: preconditioned-lag-tm-synce}. The preconditioned framework is obtained by training an offline lower triangular transport map from 100,000 samples of the banana distribution and then using it to precondition the MCMC sampling of the quartic distribution. As shown in~\Cref{fig: precondition-lag}, the autocorrelation plot indicates poor mixing properties with an effective sample size of only 139, primarily because the banana distribution is not a good approximation of the quartic distribution. In contrast, as shown in~\Cref{fig: tm-synce-lag}, our TM-SYNCE framework acts as a better sampler even if we do not consider variance reduction. The effective sample size is 17393, demonstrating the effectiveness of transport maps in capturing the geometry of the target distribution and achieving good mixing properties even when the coarse and fine distributions are dissimilar.

\subsection{Spalart--Allmaras turbulence model}\label{subsec: spalart--allmaras}
We now present results of our proposed transport map integrated SYNCE framework on a more complex problem of turbulence parameter estimation using the Spalart--Allmaras turbulence model~\cite{spalartOneequationTurbulenceModel}. The inverse problem involves estimating the turbulence model parameters $\bm{\theta}_{SA} = [\kappa, C_{b1},\sigma_{SA}, C_{\nu1}]$ using experimental data for the NACA~0012 airfoil at a Reynolds number $Re = 6 \times 10^6$ and a mach number $M = 0.15$. In particular, we use the Ladson data~\cite{ladson1988effects,ladson_naca0012_data} lift coefficient ($C_L$) at high angles of attack ranging from $10^\circ$ to $20^\circ$ as the experimental data and assume a Gaussian noise model for the data with noise covariance 0.001. We choose the output function $Q$ also as the lift coefficient $C_L$ defined at the chosen experimental angles of attack. Note that we include high angles of attack including stall angles in our calibration process to make the problem more challenging and test the model's predictive capabilities in nonlinear flow separation regimes. We hope that this comprehensive parameter calibration approach leads to uncertainty estimates that better encompass the experimental data across both attached and separated flow.

We use two levels of fidelity in this problem --- a high-fidelity RANS solver ADflow and a low-fidelity Gaussian Process Regression model~\cite{GPflow2017} trained on 3000 high-fidelity data. ADflow solves Euler, laminar Navier--Stokes, and RANS equations using a second-order finite volume method for both multi-block structured and overset meshes~\cite{Mader2020a}. The original Newton--Krylov solver implementation has been superseded by the implementation of a more robust approximate approximate Newton--Krylov algorithm~\cite{Yildirim2019b}. ADflow has several turbulence models available, including SA~\cite{spalartOneequationTurbulenceModel}, $k-\omega$~\cite{wilcoxFormulationKwTurbulence2008} and the Menter shear stress transport~\cite{menterTwoequationEddyviscosityTurbulence1994}. The default turbulence model in ADflow is the SA model, which is fully differentiated for gradient computations with the adjoint method~\cite{Kenway2019a}. ADflow can use different variants of the SA model as defined in NASA Turbulence Modeling Resource~\cite{TurbulenceModelingResource}. In this work, we use the SA-noft2 variant. 

For brevity, we do not describe the SA-noft2 model in detail. The model can be represented as 
\begin{align}\label{eq: SA}
    \frac{\partial \tilde{\nu}}{\partial t} + \mathbf{u}_j \frac{\partial \tilde{\nu}}{\partial x_j} = P(\tilde{\nu}, \bm{\theta}) - M(\tilde{\nu}, \bm{\theta}) + D(\tilde{\nu}, \bm{\theta}), 
\end{align}
where $\tilde{\nu}$ is the working variable of the turbulence model, $P(\tilde{\nu})$ is the production term, $M(\tilde{\nu})$ is the destruction term, and $D(\tilde{\nu})$ is the diffusion term and $\bm{\theta}$ represents the model parameters. In the original formulation~\cite{spalartOneequationTurbulenceModel}, the model includes 11 closure coefficients, but the SA-noft2 variant ignores the $f_{t2}$ term and reduces the number of coefficients to 10. As suggested by~\cite{schaeferUncertaintyQuantificationSensitivity2017,maruyamaDatadrivenBayesianInference2021,daronchSensitivityCalibrationTurbulence2020}, we reduce the number of coefficients even further to 4 to benefit computational costs and avoid non-identifiability issues. The four coefficients are $\kappa$ that influences the near wall behavior, $C_{b1}$ that scales the production term, $\sigma_{SA}$ that controls the diffusion term, and, $C_{\nu1}$ that scales the destruction term. These four parameters have the highest Sobol indices for majority of the test cases experimented in~\cite{schaeferUncertaintyQuantificationSensitivity2017}. The choice of the epistemic intervals that serve as priors for the four coefficients is taken from~\cite{maruyamaDatadrivenBayesianInference2021,daronchSensitivityCalibrationTurbulence2020} and is shown in~\Cref{tab: SA_coeffs}. The intervals are chosen to be 0.5 and 1.5 times the nominal values of the coefficients. The nominal values are taken from~\cite{TurbulenceModelingResource} and are also shown in~\Cref{tab: SA_coeffs}. 

\begin{table}[htbp]
    \centering
    \caption{Nominal values and bounds for the SA model coefficients}
    \label{tab: SA_coeffs}
    \begin{tabular}{lccc}
        \hline
        coefficient & nominal value & lower bound & upper bound \\
        \hline
        $\kappa$ & 0.41 & 0.205 & 0.615 \\
        $C_{b1}$ & 0.1355 & 0.06775 & 0.20325 \\
        $\sigma_{SA}$ & 0.6667 & 0.33335 & 1.0005 \\
        $C_{\nu1}$ & 7.1 & 3.55 & 10.65 \\
        \hline
    \end{tabular}
\end{table}

We run the TM-SYNCE coupling~\Cref{alg:tm-synce-coupling} with lower triangular transport maps of total order 3 for both the coarse and fine levels. For the reference proposal, we use a Gaussian distribution with an adaptive covariance defined by Haario's algorithm~\cite{haarioAdaptiveMetropolisAlgorithm2001} and set the resynchronization parameter $\omega=0.5$. We run 5 independent times for 10000 samples and discard the first 2000 samples as burn-in. We adapt the transport map using~\Cref{alg:optimization} and set $K = 250$. 

\begin{figure}[ht]
    \centering
    \includegraphics[width=0.8\textwidth]{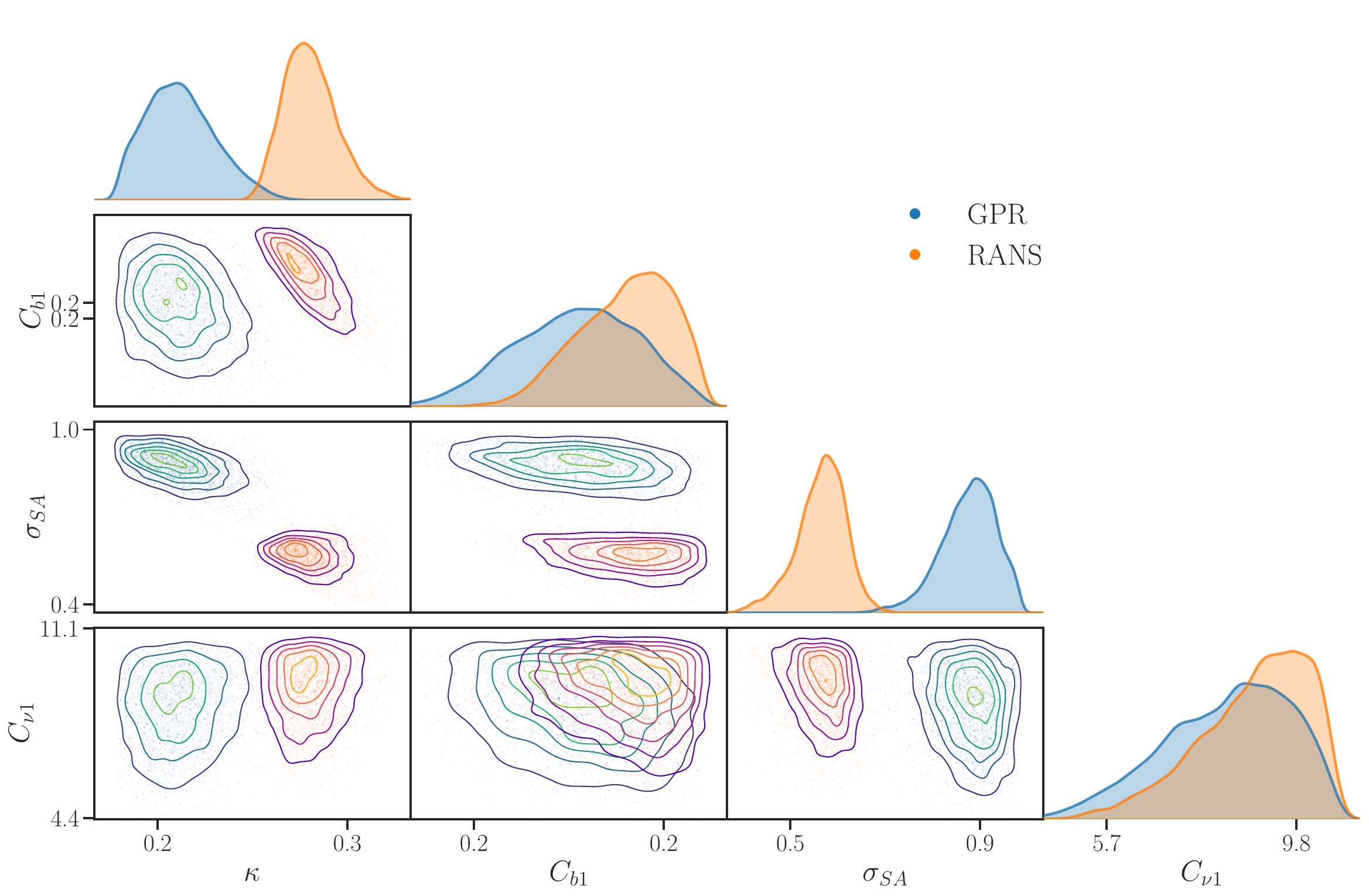}
    \caption{Corner plot showing the marginal distribution on the diagonal and the joints between all the turbulence model parameters below the diagonal. Existing multi-fidelity methods fail to generate correlated samples when the posteriors are dissimilar as shown here. Achieving good sampling efficiency is also difficult because of the complex geometry of the posteriors.}
    \label{fig: naca-scatter}
\end{figure}

\begin{figure}[ht]
    \centering
    \begin{subfigure}[b]{0.24\textwidth}
        \centering
        \def\svgwidth{\textwidth}
        \includegraphics[width=\textwidth]{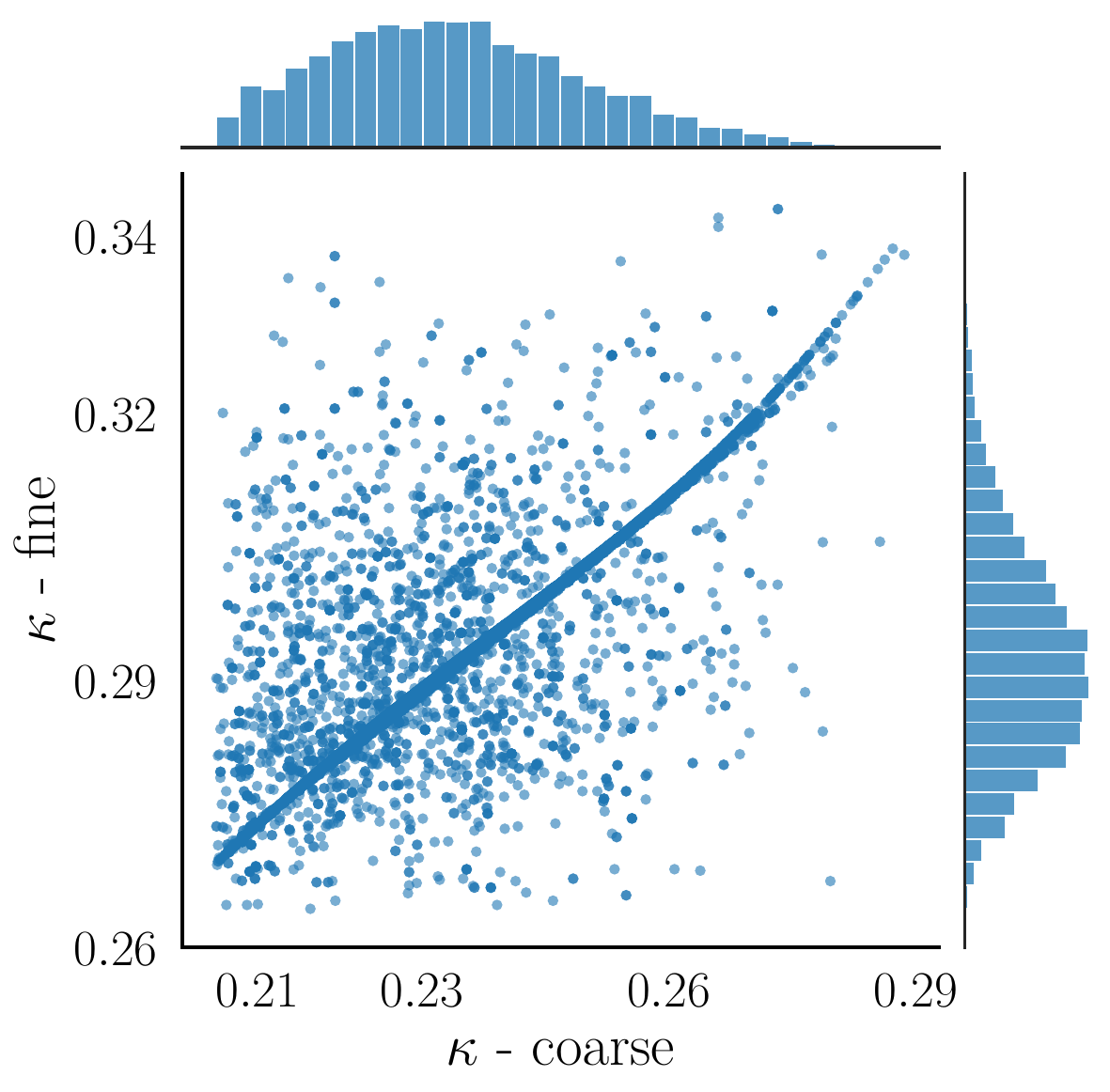}
        \caption{$\rho = 0.75$}
        \label{fig: dim-1}
    \end{subfigure}
    \hfill
    \begin{subfigure}[b]{0.24\textwidth}
        \centering
        \def\svgwidth{\textwidth}
        \includegraphics[width=\textwidth]{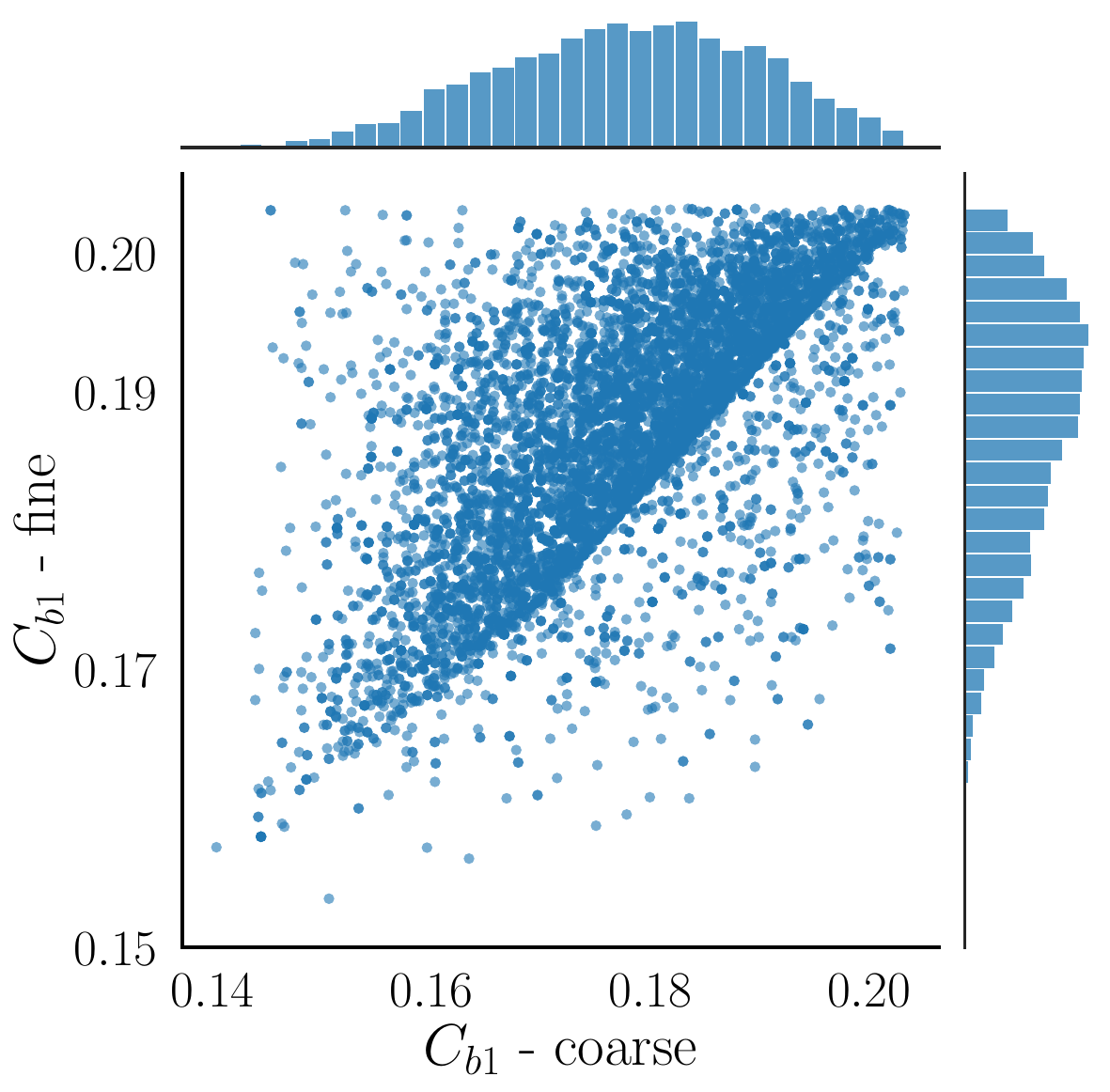}
        \caption{$\rho = 0.66$}
        \label{fig: dim-2}
    \end{subfigure}
    \hfill
    \begin{subfigure}[b]{0.24\textwidth}
        \centering
        \def\svgwidth{\textwidth}
        \includegraphics[width=\textwidth]{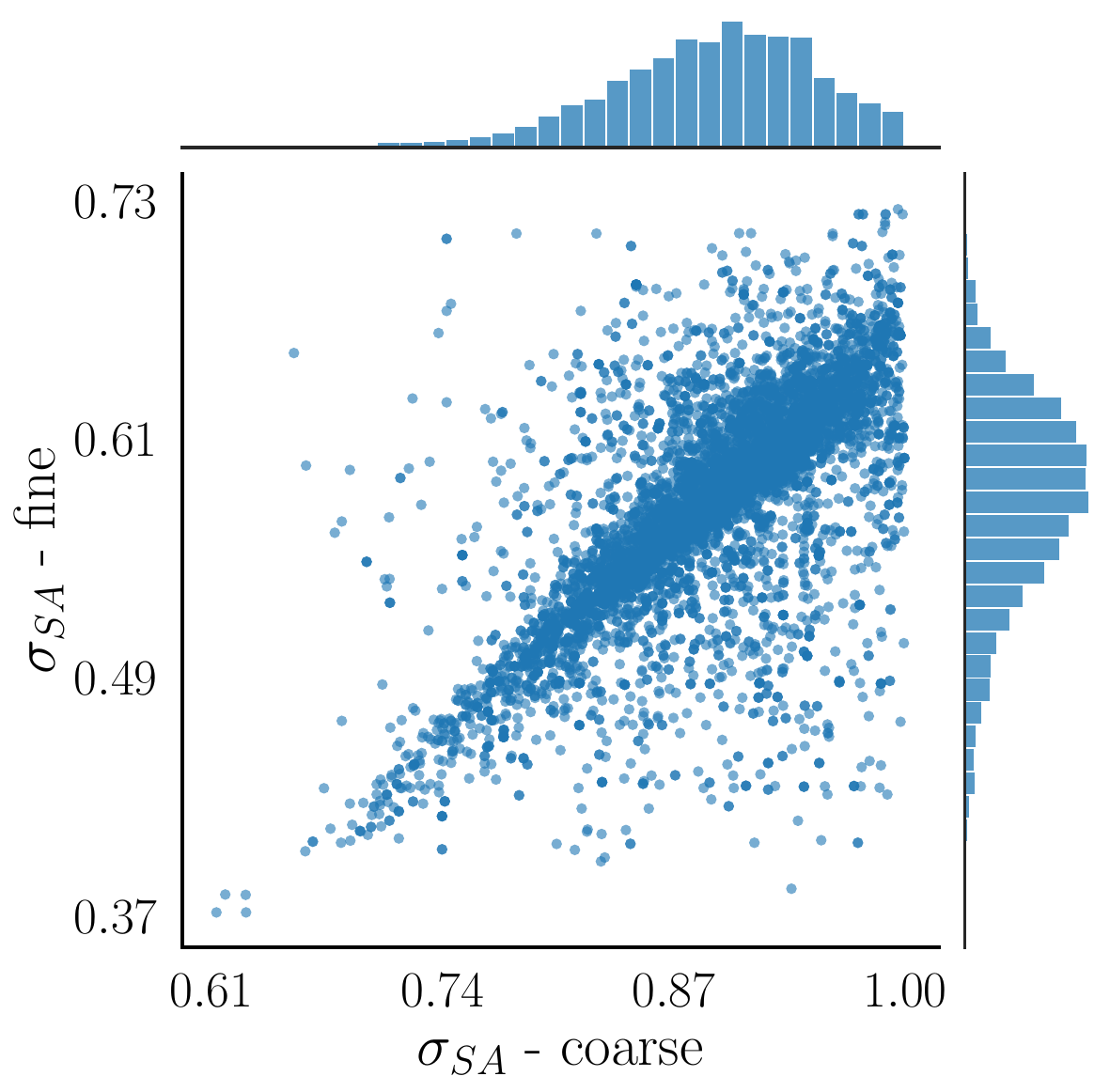}
        \caption{$\rho = 0.69$}
        \label{fig: dim-3}
    \end{subfigure}
    \hfill
    \begin{subfigure}[b]{0.24\textwidth}
        \centering
        \def\svgwidth{\textwidth}
        \includegraphics[width=\textwidth]{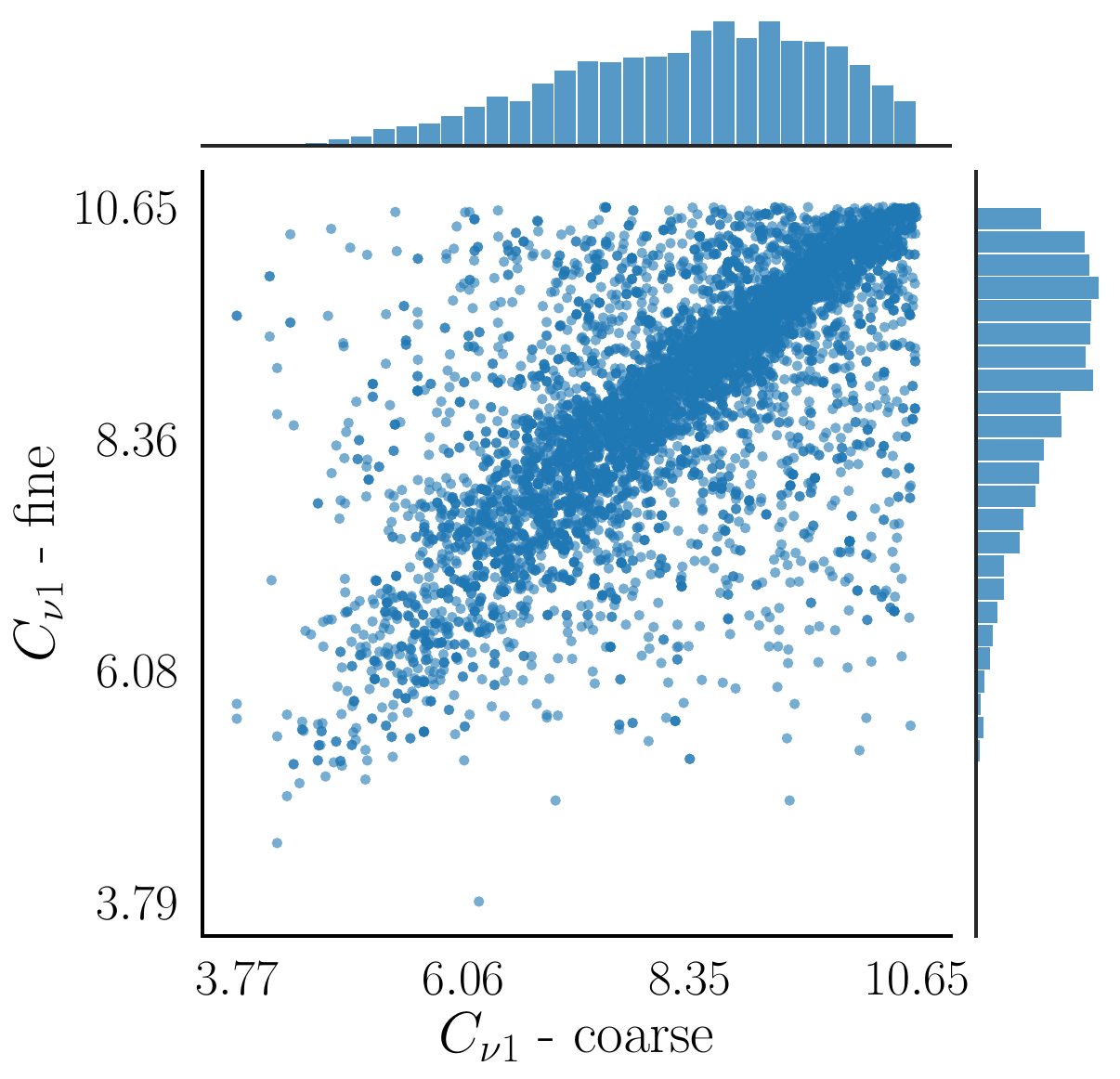}
        \caption{$\rho = 0.72$}
        \label{fig: dim-4}
    \end{subfigure}
    \caption{Joint scatter plots of the turbulence model parameters obtained from the TM-SYNCE coupling algorithm. More samples falling on the diagonal indicate better correlation between the samples as the Pearson correlation coefficient measures linear correlation.}
    \label{fig: joint-plots}
\end{figure}

\begin{figure}[ht]
    \centering
    \begin{subfigure}[b]{0.48\textwidth}
        \centering
        \def\svgwidth{\textwidth}
        \includegraphics[width=\textwidth]{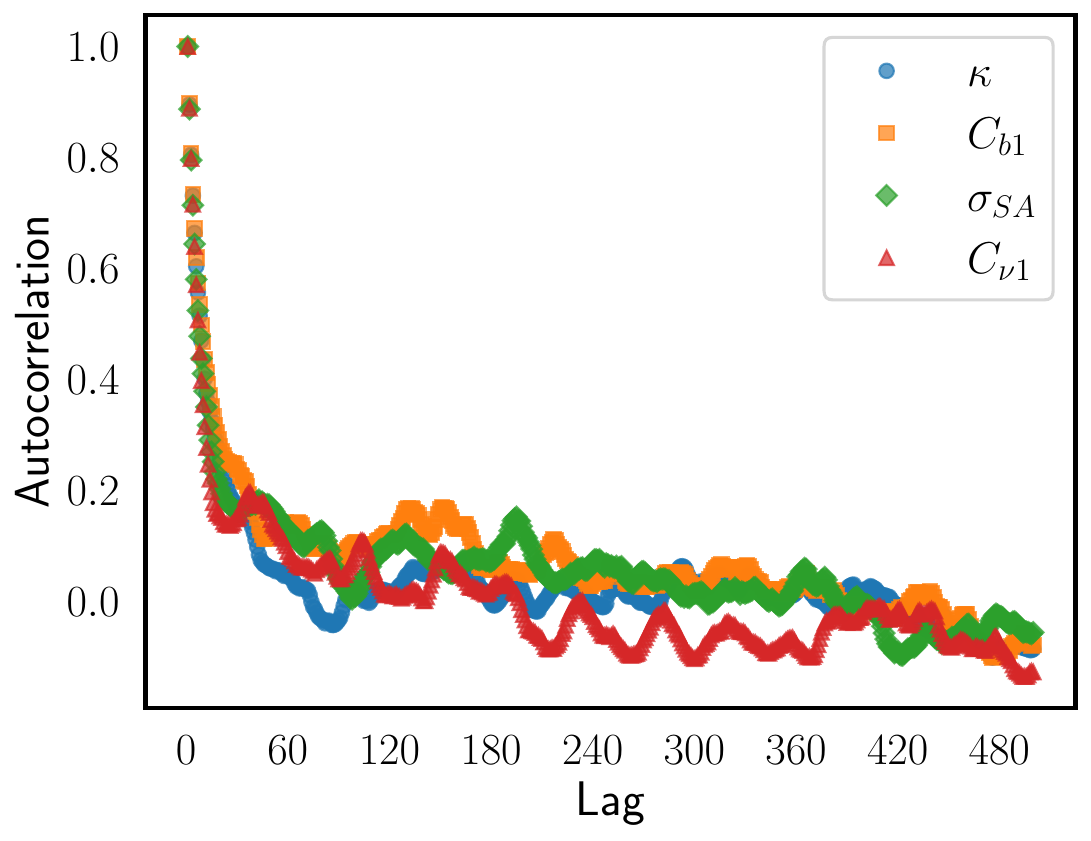}
        \caption{Single fidelity MCMC without transport maps. Effective sample size is 28.}
        \label{fig: naca-lag-single}
    \end{subfigure}
    \hfill
    \begin{subfigure}[b]{0.48\textwidth}
        \centering
        \def\svgwidth{\textwidth}
        \includegraphics[width=\textwidth]{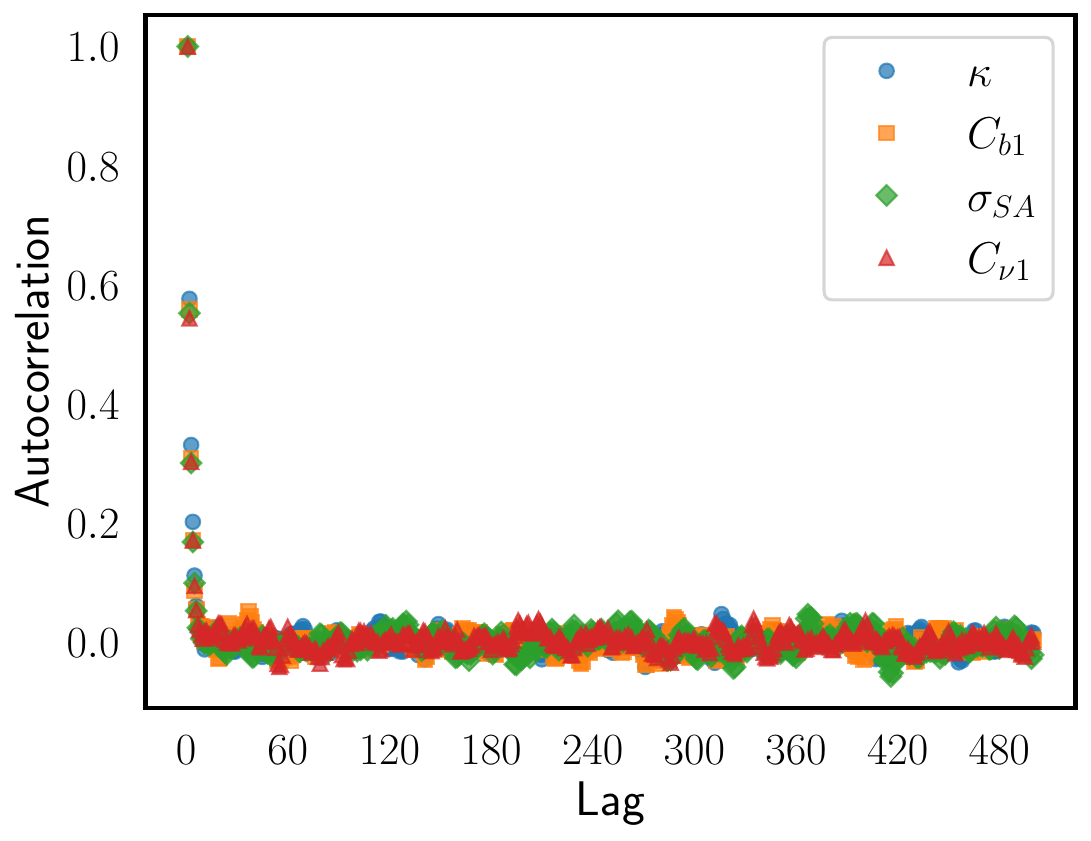}
        \caption{Bi-fidelity MCMC with LT transport map~\cite{parnoTransportMapAccelerated2018}. Effective sample size is 2075.}
        \label{fig: naca-lag-transport}
    \end{subfigure}
    \caption{Autocorrelation plots for the high-fidelity RANS samples comparing with and without transport maps. We see huge improvements in the sampling efficiency because of the transport maps.}
    \label{fig: naca-lag}
\end{figure}

Figure~\ref{fig: naca-scatter} shows how non Gaussian and different the two posteriors are with respect to each other rendering correlation and sampling difficult. The different posteriors arise because the low-fidelity GPR model is trained on a limited number of high-fidelity samples and does not capture the full complexity of the high-fidelity RANS model.
In~\Cref{fig: joint-plots}, we show the joint scatter plots of the turbulence model parameters obtained from the TM-SYNCE framework. These plots show how well the samples are correlated between the two posteriors. The more samples falling on the diagonal line, the better the correlation between the samples. In~\Cref{fig: naca-lag}, we show the autocorrelation versus lag plots for the high-fidelity samples obtained by single-fidelity MCMC without transport maps and bi-fidelity MCMC with transport maps. Haario's covariance adaptation~\cite{haarioAdaptiveMetropolisAlgorithm2001} is used in both cases to ensure a fair comparison. In the single fidelity without transport maps case, adaptation is performed directly on the target space, while in the bi-fidelity with transport maps case, adaptation is performed in the reference space. The decay in autocorrelation is steep in the transport map case leading to low integrated autocorrelation time and high effective sample size. These results indicate that the transport maps are able to capture the geometry of the posteriors and supplement SYNCE to generate correlated samples that are well mixed.

\begin{figure}
    \centering
    \begin{subfigure}[b]{0.48\textwidth}
        \centering
        \def\svgwidth{\textwidth}
        \includegraphics[width=\textwidth]{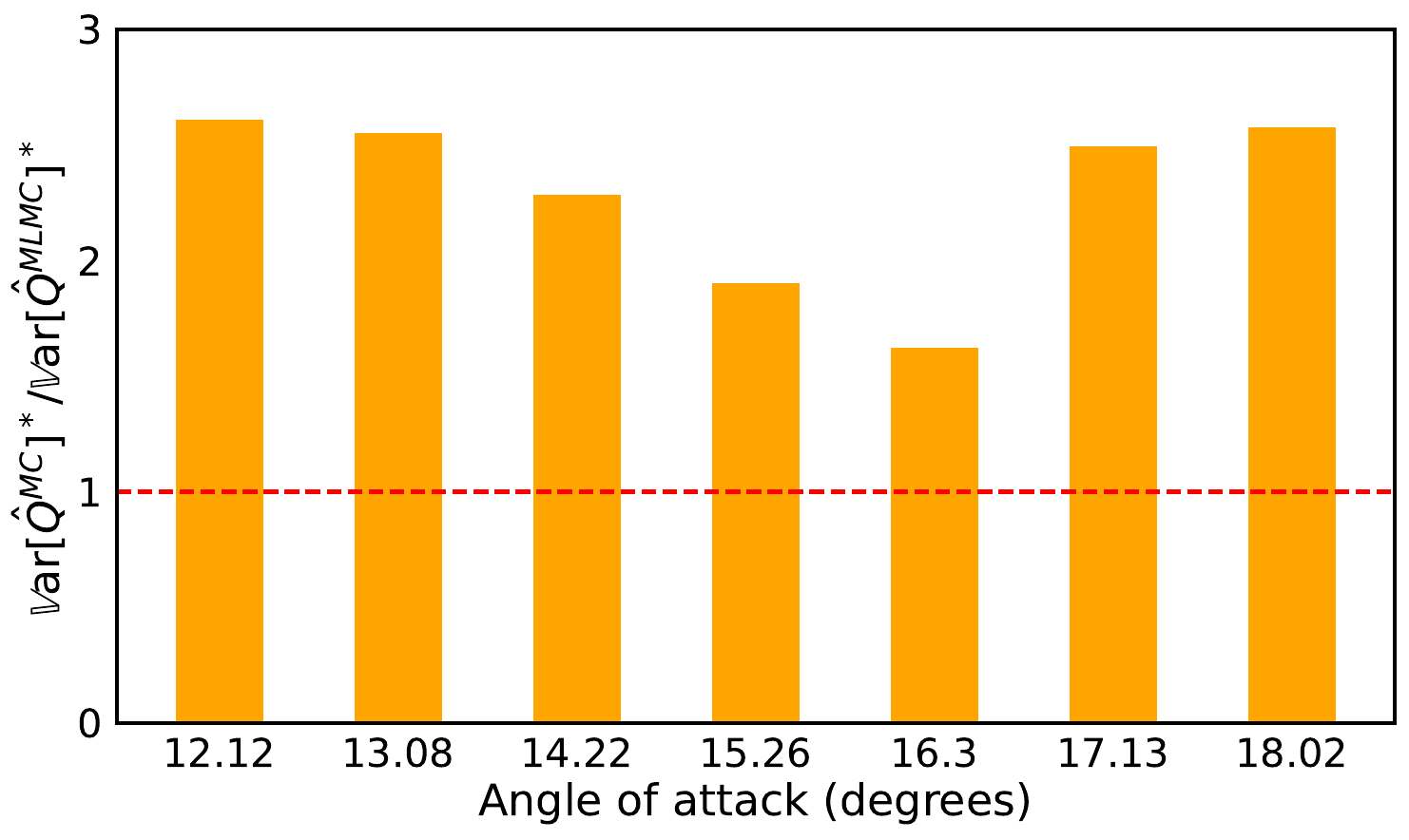}
        \caption{Variance ratio plot between high-fidelity Monte Carlo estimator and the multi-level Monte Carlo estimator.}
        \label{fig: naca-variance-reduction}
    \end{subfigure}
    \hfill
    \begin{subfigure}[b]{0.48\textwidth}
        \centering
        \def\svgwidth{\textwidth}
        \includegraphics[width=\textwidth]{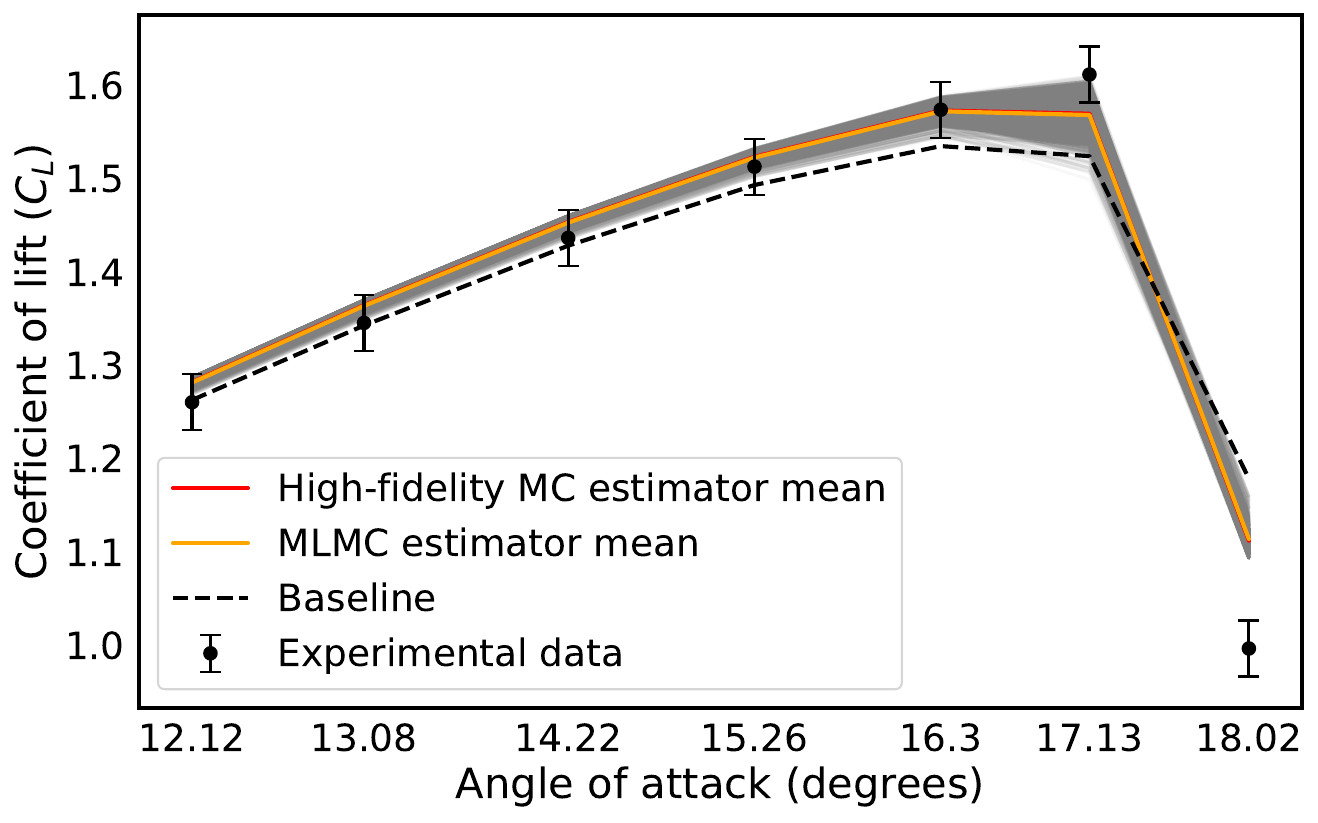}
        \caption{Posterior predictive polar plots with samples obtained from the TM-SYNCE coupling framework.}
        \label{fig: naca-prediction}
    \end{subfigure}
    \caption{Variance ratio and polar plot for the NACA0012 SA turbulence model parameter estimation problem.}
    \label{fig: naca-variance-prediction}
\end{figure}

The improvements in correlation and mixing properties translate to significant computational cost savings through the multi-fidelity MLMC framework. In~\Cref{fig: naca-variance-reduction}, we plot the variance ratio between multi-level Monte Carlo (MLMC) estimator that uses optimal sample allocation~\cite{gilesMultilevelMonteCarlo2015} and an equivalent cost high-fidelity Monte Carlo (MC) estimator by assuming a cost ratio of 0.001 between the low-fidelity and high-fidelity models. We see close to $2.5\times$ variance reduction over all angles of attack translating to $50\%$ time savings for models that have dissimilar posteriors, a promising result indicating that our method works for any low-fidelity model, not just models that are \textit{similar} to the high-fidelity model. See Appendix~\ref{sec: appendix} for a detailed derivation of the optimal sample allocation, variance ratio, and their impact on cost savings.

In~\Cref{fig: naca-prediction}, we plot the output functional mean for the high-fidelity MC and bi-fidelity MLMC estimator along with predictions from some random samples obtained using the coupling framework. We accomplish the goal of calibrating the parameters to better match experimental data, especially at high angles of attack near stall regions where UQ estimates are notoriously expensive to find because of the computational costs involved. The high-fidelity MC estimator mean and the MLMC estimator mean are close to each other, indicating that the MLMC estimator is unbiased w.r.t the high-fidelity model but with significantly lower variance. 

Despite the significant variance reduction and improved predictive capabilities achieved through calibration of the SA turbulence model parameters via our TM-SYNCE framework, we observe persistent discrepancies between calibrated model predictions and experimental data at high angles of attack. This underscores that these differences are not merely due to parameter uncertainty but rather stem from fundamental limitations in the Spalart-Allmaras (SA) turbulence model itself. The SA model, while robust and computationally efficient for many aerodynamic flows, is a relatively simple, single-equation eddy-viscosity model based on the Boussinesq hypothesis. It struggles to accurately capture complex flow physics in nonlinear regimes such as flow separation, rapid strain rate changes, and stall, where flow anisotropy, unsteady effects, and nonlinear turbulence interactions become dominant. These inherent inadequacies in the model form manifest as systematic errors in predictions that cannot be fully corrected through parameter tuning.

Therefore, the discrepancies reflect model form error —- errors arising from assumptions and simplifications embedded in the turbulence closure formulation. Addressing these errors requires moving beyond parameter calibration to incorporate more sophisticated turbulence models that can represent nonlinear flow features with higher fidelity. Examples include Reynolds stress transport models or data-driven turbulence closures that explicitly model anisotropy and nonlinear effects. Future work will focus on quantifying and reducing model form errors through model error quantification frameworks and model discrepancy approaches. By explicitly characterizing model inadequacy, it may be possible to separate uncertainty due to parameters from structural model limitations and systematically improve predictive accuracy, especially in challenging nonlinear flow regimes like stall.

\section{Conclusion}\label{sec: conclusion}
In this work, we developed a novel framework that addresses a fundamental challenge in multi-fidelity inverse problems, specifically the difficulty of generating correlated samples between dissimilar high- and low-fidelity models while maintaining good sampling efficiency. Our transport map integrated SYNCE framework provides a systematic approach to couple samples from different fidelity levels with arbitrary posterior distributions, enabling efficient sampling and variance reduction in output estimates.

The key idea of the proposed framework is to leverage transport maps to create a common reference space where proposals can be easily generated and correlated. SYNCE provides the coupling mechanism in the reference space where both random walk and independent proposals are controlled by a resynchronization parameter. The coupling method itself is simple and uses a common random number in the reference space to create coupled proposals, which are then mapped to the target space through transport maps. We presented the theoretical advantages of using optimal transport maps and their practical alternatives, including lower triangular transport maps and normalizing flows. We also introduced the notion of direct and deep configurations for transport maps and discussed their trade-offs in terms of computational efficiency and accuracy. We provided an adaptive algorithm for transport map optimization during MCMC sampling that ensures that the maps remain accurate as more samples are collected.

We then demonstrated the effectiveness of our framework on a simple test problem with banana and quartic distributions, as well as a more complex turbulence parameter estimation problem using the Spalart--Allmaras turbulence model on the NACA~0012 airfoil. The results for the first test problem compared the performance of different transport map architectures and adaptation strategies, showing that our framework achieves significant improvements in correlation, effective sample size, and mixing properties compared to traditional single-fidelity and existing multi-fidelity sampling methods. The second test problem demonstrated the practical applicability of our framework in a real-world turbulence parameter estimation problem, achieving significant variance reduction in output estimates and leading to substantial cost savings in obtaining these estimates. Our framework reduced the variance of output estimates over all angles of attack by a factor of almost $2.5\times$, translating to $50\%$ time savings compared to high-fidelity only approaches.

A crucial aspect of our framework is its applicability to any low-fidelity model, not just models that are similar to the high-fidelity model. This is particularly important in real-world applications where data-driven black-box models are often used as low-fidelity surrogates. The ability to generate correlated samples between dissimilar posteriors opens up new avenues for uncertainty quantification and parameter estimation in complex engineering systems where model hierarchies exhibit significant differences.

\section*{Appendix}\label{sec: appendix}
The variance of a bi-fidelity MLMC estimator is given by
\begin{align*}
    \mathbb{V}\text{ar}&\left[\hat{Q}^{\text{MLMC}}_1\right] = \frac{V_0}{N_0} + \frac{V_1}{N_1},
\end{align*}
where $V_0 = \mathbb{V}\text{ar}\left[Q_0\right]$ and $V_1 = \mathbb{V}\text{ar}\left[Q_1 - Q_0\right]$ is the coupled variance of the fine and coarse levels. Here, $Q_0$ and $Q_1$ are the coarse and fine level output functionals, respectively. For our turbulence parameter estimation problem, $Q_0$ is the lift coefficient $C_L$ vector computed using the low-fidelity GPR model and $Q_1$ is the lift coefficient $C_L$ vector computed using the high-fidelity RANS model at the experimentally defined angles of attack. 

To minimize the computational cost of the MLMC estimator for a fixed target error $\epsilon^2$, we minimize the total cost,
\begin{align*}
    C = N_0 C_0 + N_1 C_1,
\end{align*}
subject to the constraint $\sum \mathbb{V}\text{ar}\left[\hat{Q}^{\text{MLMC}}_1\right] = \epsilon^2 / 2$, where $C_0$ is the computational cost per coarse model evaluation and $C_1$ is the computational cost per coupled sample (both coarse and fine model evaluations). This optimization problem can be solved using the method of Lagrange multipliers~\cite{gilesMultilevelMonteCarlo2015}. The optimal sample sizes $N_0^*$ and $N_1^*$ are given by,
\begin{align*}
    N_0^* &= 2 \epsilon^{-2} \sqrt{\frac{\tilde{V}_0}{C_0}} \left(\sqrt{\tilde{V}_0C_0 + \tilde{V}_1C_1}\right) \\
    N_1^* &= 2 \epsilon^{-2} \sqrt{\frac{\tilde{V}_1}{C_1}} \left(\sqrt{\tilde{V}_0C_0 + \tilde{V}_1C_1}\right),
\end{align*}
where $\tilde{V}_0$ is the sum of variances of the $V_0$ vector and $\tilde{V}_1$ is the sum of variances of $V_1$ vector~\cite{thomasgroupedACV2024}. The sum represents the trace of the covariance matrix of the output functional at all angles of attack.

The total cost of the optimally allocated MLMC estimator is given by
\begin{align*}
    C^* &= N_0^* C_0 + N_1^* C_1 \\
    &= 2 \epsilon^{-2} \left(\sqrt{\tilde{V}_0C_0 + \tilde{V}_1C_1}\right) \left(\sqrt{\tilde{V}_0C_0} + \sqrt{\tilde{V}_1C_1}\right),
\end{align*}
and the variance of the optimally allocated MLMC estimator is given by
\begin{align*}
    \mathbb{V}\text{ar}\left[\hat{Q}^{\text{MLMC}}_1\right]^* &= \frac{V_0}{N_0^*} + \frac{V_1}{N_1^*} \\
    &= \frac{1}{2 \epsilon^{-2} \left(\sqrt{\tilde{V}_0C_0 + \tilde{V}_1C_1}\right)}\left[V_0\sqrt{\frac{C_0}{\tilde{V}_0}} + V_1\sqrt{\frac{C_1}{\tilde{V}_1}}\right].
\end{align*}
The number of equivalent cost high-fidelity samples $N_{\text{eq}}$ that can be obtained for the same cost as the optimally allocated MLMC estimator is given by
\begin{align*}
    N_{\text{eq}} &= \frac{C^*}{C_1 - C_0} \\
     &= \frac{2 \epsilon^{-2} \left(\sqrt{\tilde{V}_0C_0 + \tilde{V}_1C_1}\right) \left(\sqrt{\tilde{V}_0C_0} + \sqrt{\tilde{V}_1C_1}\right)}{\left(C_1 - C_0\right)},
\end{align*}
and the variance of the equivalent cost high-fidelity MC estimator is given by
\begin{align*}
    \mathbb{V}\text{ar}\left[\hat{Q}^{\text{MC}}_1\right]^* &= \frac{\mathbb{V}\text{ar}\left[Q_1\right]}{N_{\text{eq}}}.
\end{align*}
Finally, the ratio of the optimal sample allocated MLMC variance to the equivalent cost high-fidelity MC estimator variance is given by
\begin{align*}
    \frac{\mathbb{V}\text{ar}\left[\hat{Q}^{\text{MC}}_1\right]^*}{\mathbb{V}\text{ar}\left[\hat{Q}^{\text{MLMC}}_1\right]^*} = \frac{\mathbb{V}\text{ar}\left[Q_1\right] (C_1 - C_0)}{\left(\sqrt{\tilde{V}_0C_0} + \sqrt{\tilde{V}_1C_1}\right)\left[V_0\sqrt{\frac{C_0}{\tilde{V}_0}} + V_1\sqrt{\frac{C_1}{\tilde{V}_1}}\right]}.
\end{align*}
Having a computationally inexpensive low-fidelity model ($C_0 \ll C_1$) that produces output quantities highly correlated with the high-fidelity results ($V_1 \ll V_0$) leads to small denominator values in the above equation, resulting in a large variance ratio and significant computational savings. 

Instead of comparing the variances, we can also compare the costs between the optimally allocated MLMC estimator and a high-fidelity single level estimator that achieves the target variance $\epsilon^2 / 2$. For the high-fidelity single level estimator to match the target variance, we need to use
\begin{align*}
    N_{\text{target}} &= \frac{\sum \mathbb{V}\text{ar}\left[\hat{Q}^{\text{MC}}_1\right]}{\epsilon^2 / 2}.
\end{align*}
We can now compare the cost of this high-fidelity single level MC estimator $N_{\text{target}} \left(C_1 - C_0\right)$ to the cost of the optimally allocated MLMC estimator $C^*$.

\section*{Code Availability}
The complete source code and scripts used to run all experiments reported in this paper are publicly available at the following GitHub repository: \href{https://github.com/sanjan98/MF-TM-SYNCE}{https://github.com/sanjan98/MF-TM-SYNCE}.

\section*{Funding Sources}

This work was funded by the Aeronautics Research Mission Directorate at NASA through the Transformative Aeronautical Concepts Program (TACP) and the D.2 Transformational Tools and Technologies Project (TTT) under contract no.~80NSSC23M0215.

\bibliography{references}

\end{document}